\documentclass[%
  reprint,
 superscriptaddress,
bibnotes, 
 amsmath,amssymb,
 aps,
pra,
]{revtex4-2}
\usepackage{color}

\usepackage{lmodern} 
\usepackage{orcidlink}
\usepackage{graphicx}
\usepackage{physics}
\usepackage{xspace}
\newcommand{\ps}{phase space\xspace}
\newcommand{\cat}{`cat'\xspace}
\newcommand{\squ}{{s}}
\newcommand{\state}{{{\varrho}}}
\newcommand{\cti}{{universal}\xspace}

\newcommand{\sens}{\it sensi\-tive\xspace}
\usepackage{hyperref}
\usepackage{soul}
\usepackage[normalem]{ulem}
\newcommand{\VEC}[1]{{\mbox{\boldmath${#1}$}}}

\makeatletter
\newsavebox{\@brx}
\newcommand{\llangle}[1][]{\savebox{\@brx}{\(\m@th{#1\langle}\)}%
  \mathopen{\copy\@brx\kern-0.5\wd\@brx\usebox{\@brx}}}
\newcommand{\rrangle}[1][]{\savebox{\@brx}{\(\m@th{#1\rangle}\)}%
  \mathclose{\copy\@brx\kern-0.5\wd\@brx\usebox{\@brx}}}
\makeatother

\usepackage{forloop,ifthen} 

\usepackage{xifthen}
\newcommand{\refAppendix}[6]{#1
  \ifthenelse{\isempty{#2}}%
    {}
    {\protect\cite{#2}}
    #3\protect\ref{#4}#5#6\xspace
}
\usepackage{tcolorbox}
\definecolor{skyblue}{rgb}{0.53, 0.81, 0.92}

\begin{document}

\title{A Sensitive Quantumness Measure for One-Dimensional Continuous-Variable Systems}

\author{Ole Steuernagel\orcidlink{0000-0001-6089-7022}}
\email{Ole.Steuernagel@gmail.com}
\affiliation{Institute of Photonics Technologies, National Tsing Hua University, Hsinchu 30013, Taiwan}

\author{Hsien-Yi Hsieh\orcidlink{0000-0001-5227-8248}}
\affiliation{Institute of Photonics Technologies, National Tsing Hua University, Hsinchu 30013, Taiwan}

\author{Yi-Ru Chen\orcidlink{0000-0001-8580-9025}} 
\affiliation{Institute of Photonics Technologies, National Tsing Hua University, Hsinchu 30013, Taiwan}

\author{Ray-Kuang Lee\orcidlink{0000-0002-7171-7274}}
\affiliation{Institute of Photonics Technologies, National Tsing Hua University, Hsinchu 30013, Taiwan}
\affiliation{Department of Physics, National Tsing Hua University, Hsinchu 30013, Taiwan}
\affiliation{Physics Division, National Center for Theoretical Sciences, Taipei 10617, Taiwan}
\affiliation{Center for Quantum Science and Technology, Hsinchu, 30013, Taiwan}
 
\date{\today}
\begin{abstract}
  For one-dimensional continuous-variable quantum systems such as single-mode quantum optical
  systems, we give a quantification of the quantumness of such a system's state,~$\state$, by
  introducing the \emph{measure of quantumness},~$\Xi$, which works for all states, pure or
  mixed. $\Xi$~is a measure which is \emph{\cti, sensitive}, \emph{mono\-tonic}, and
  \emph{unbounded}. $\Xi[\state]$~yields a single positive value to quantify how
  nonclassical~$\state$ is.  $\Xi$~employs~\ps distri\-butions to represent~$\state$ and is a fixed
  function,~$\Xi[{\boldsymbol{\cdot}}]$, independent of the system, its environment or the type of
  state.
\end{abstract}

\maketitle

\section{Introduction and Motivation}

So far, it is not known how to universally quantify quantumness in general physical
systems~\cite{Wood__Quantumness_Quanta23}. Even for continu\-ous systems of one dimension, whose
pure states are described by single-argument wavefunctions,~$\psi(x)$, this is not
known~\cite{Froewis_RMP18}.

But, for such 1-D systems~\cite{Titulaer_Glauber__PR66}, Bohmann and Agudelo~\cite{Bohmann_PRL20}
recently devised a quantumness certification functiona\textsl{}l which is very sensitive, we denote
it by $\xi(x,p)$. It is based in \ps, $x$ is position and $p$ its associated momentum, and built
from Wigner's, $W(x,p)$, and Husimi's, $Q(x,p)$,
distributions~\cite{Cahill_PR69b,Schleich_01,Scully_Zubairy__Book01}.

Our main interest in the quantumness certification functional $\xi$ results from the following
considerations:

We deal with experimentally generated impure squeezed states, for these $\xi(x,p)$ is
\emph{faithfully discriminating}, it is sensitive and, for all gaussian
states~\cite{Ole_25_TowardsQuantumness} can flawlessly tell quantum and classical states
apart~\cite{Wuensche_JOBQSO04}. Our states are impure and we want to make sure that they are not
mislabelled as classical when showing fluctuations below the level of vacuum state
fluctuations~\cite{Wuensche_JOBQSO04}.

Additionally, the measure of quantumness~$\Xi[\state]$ which we build from $\xi(x,p)$ scales
quadratically with the size of a \cat state, as, in the light of
Ref.~\cite{Nimmrichter_Hornberger_PRL13}, it should. The measure~$\Xi$ presented here, is not only
conceptually simple to use but, as far as we can tell, also the most
sensitive~\cite{Ole_25_TowardsQuantumness} available so far, see
Appendix~\ref{subsec:OtherMeasures}.

We will next introduce some further basic considerations, then highlight $\Xi$'s fundamental
features, we demand from it, in Sections~\ref{sec:_Xi_Universal_Discriminating}
and~\ref{sec:_Xi_Monotonic_Unbounded}, introduce \ps distributions in
Sect.~\ref{sec:Liouville_W_dist}, show that $\Xi$ grows with increasing coherent spread in \ps, in
Sections~\ref{sec:_Xi_grows_coherence},~\ref{sec:_Xi_grows_quadratically},~\ref{sec:_Xi_grows_unbound},
and~\ref{sec:_Xi_grows_squeezedStates}. Next we show that $\Xi$ is convex when quantifying mixed
states in Sect.~\ref{sec:_mixed_states} and apply it to our experimental data in
Sect.~\ref{sec:_apply_experimentatl_data}, some further general considerations of features of $\Xi$
are considered in the remaining Sections and Appendices, together with our
Conclusions~\ref{sec:Conclusion}.

\subsection*{Quantumness Certification Functional $\xi$}
Since the Wigner distribution $W$ is the Fourier-transform~(\ref{eq:_WignerDistr}) of $\state$, we
use $\state$ and $W$ inter\-changeably to denote states, with this proviso: $\xi[\state]$'s explicit
form is
\begin{flalign}
  & \xi[W](x,p) =  W(x,p) - 4 \pi \; Q^2(x,p) \; ,
  \label{eq:xi_Certification} 
\end{flalign}
$\xi$ is the most sensitive certification functional we could find: $\xi[W](x,p) < 0 $ certifies
that a state~$W$ is nonclassical; if $W$ is classical, then $\xi \geq 0$ throughout
\ps~\cite{Bohmann_PRL20}. But there are weakly nonclassical states for which $\xi \geq 0$: $\xi$
does not always faithfully discriminate between classical and quantum
states~\cite{Ole_25_TowardsQuantumness}.

$\xi$ is always mathema\-ti\-cally well beha\-ved since $W$ and $Q$ are~\cite{Cahill_PR69b}. This
opens the door for our construction of $\Xi[\state]$:
%
%
\begin{flalign}
  \Xi[W] = \iint_{-\infty}^{\infty} dx \; dp \;\; \big.\VEC{\Delta} \xi[W] (x,p)\Big|_{\xi < 0} \; ,
  \label{eq:Xi_Measure} 
\end{flalign}
where,
$\VEC{\Delta} \xi = \frac{\partial^2 \xi}{\partial x^2} + \frac{\partial^2 \xi}{\partial p^2}$ is
the \ps Laplacian and our quantum\-ness measure only includes areas in \ps (`basins' where
$\xi(x,p) <0$) where a state displays non\-classicality.

Our quantumness measure $\Xi$ contains a state's \emph{\mbox{local} quantumness},
$ \VEC{\Delta} \xi (x,p)|_{\xi < 0} $, which we will show to be approxi\-mately proportional to
coherences across \ps and to quadratically increase with the effec\-tive distan\-ces
these coherences span.

\subsection*{When is a state nonclassical? \label{subsec:_state_nonclassical}}

We follow the approach that has stood the test of time~\cite{Bohmann_PRL20}: ``the incompatibility
of the studied quantum system with classical physics is indicated through negativities in its
quasiprobability distributions. This concept has a long-standing tradition in quantum optics where
negativities in the Glauber-Sudarshan $P$ function~\cite{Sudarshan__PRL63,Glauber__PR63}, i.e., the
state expansion in terms of a statistical mixture of (classical) coherent states, forms the very
definition of nonclassicality~\cite{Titulaer_Glauber__PR65,Mandel__PS86}.''

Although the Glauber-Sudarshan distribution~$P$ by itself, in principle, successfully discriminates
between classical and quantum states, practically speaking, it does not solve our discrimination
problem, neither analytically~\cite{Cahill_PR69b} nor numerically~\cite{Lvovsky_Raymer__RMP09}. It
is hard to determine $P$; typically it is so singular that it is extremely difficult to analyze or
even construct $P$~\cite{Wuensche_JOBQSO04,Brewster_Franson__JMP18}.

This forces us to use other approaches such as the certification function~$\xi$ of
Eq.~(\ref{eq:xi_Certification}) instead.

\section{$\Xi$ is Universal and sensitive\label{sec:_Xi_Universal_Discriminating}}

Like $\xi$, $\Xi[{\boldsymbol{\cdot}}]$~(\ref{eq:Xi_Measure}) is of fixed form independent of
context such as the type of system or the type of state the system is in. $\Xi[W]$ applies to all
states of one-dimensional continuous-variable systems, irrespective of whether the state is pure or
mixed and whether it has small, large, or even macroscopic~\cite{Leggett__JPCM02} extent,
irrespective of context, hamiltonian, environment or measurement proce\-dures~\cite{Froewis_RMP18}:
this is what we mean by $\Xi$ is \emph{universal}.

This \emph{universality} implies that all states are compared like-for-like, using
$\Xi[{\boldsymbol{\cdot}}]$ applies a common metric.

We note that $\xi$ is \emph{universal}, but  by itself it cannot be used as a quantumness measure
(see~\ref{sec:xi_not_measure}).

Since we build our measure~$\Xi$ from~$\xi$, using only its nonclassical contributions
(${\xi < 0}$), $\Xi$ inherits its sensiti\-vity for discriminating between classical and nonclassical
states from~$\xi$~\cite{Bohmann_PRL20}.

Specifically, $\Xi$ can always correctly quantify the quantumness of all gaussian
states~\cite{Ole_25_TowardsQuantumness} (also see Sect.~\ref{sec:_Xi_grows_squeezedStates}).

By construction, $\Xi$ is zero for all classical
states~\cite{Bohmann_PRL20,Ole_25_TowardsQuantumness} and positive if a state is certified as
nonclassical, because $\Xi$'s integral~(\ref{eq:Xi_Measure}) over the state's local quantumness
equals line integrals, along the boundaries of the negative basins ($\xi < 0$), of the cross product
between boundary segments and local gradient $\VEC{\nabla} \xi$. As the basins are negative,
$\VEC{\nabla} \xi$ always points away from them and only positive contributions to $\Xi$ can
occur, see~\ref{sec:XiFromLineIntegrals}.

\section{$\Xi$ is Monotonic and Unbounded \label{sec:_Xi_Monotonic_Unbounded}}

$\Xi$ grows \emph{monotonically}: with every increase in quantum\-ness of a state~$\state$,
$\Xi[\state]$ grows with it. Since there is no limit in principle to the extent of the `coherent
spread' of a super\-position state, $\Xi$ should be \emph{unbounded}.

Mathematically, the quantumness of a state contributes through two aspects to the magnitude of
$\Xi$: the size of the mutual coherences between spatially separate parts of the state [given by the
value,~$\varrho$, of the density matrix], and the size of the distance between these parts that are
coherent with each other [given by the square of the distances
$y$~\cite{Nimmrichter_Hornberger_PRL13} in~$\varrho(x-y,x+y)$, compare Eq.~(\ref{eq:_WignerDistr})].
To demonstrate how~$\xi$ represent these mutual coherences and their distances, and how the form of
a state's \emph{\mbox{local} quantumness} term, $ \VEC{\Delta} \xi (x,p)|_{\xi < 0} $, in
$\Xi$~(\ref{eq:Xi_Measure}) extracts them and converts them into a \emph{monotonic} and
\emph{unbounded} \mbox{measure}, we will next introduce Wigner distri\-butions and then 
consider representative states:

Coherent state superposition or (\cat) states, spread-out (Fock) states and below-vacuum
fluctu\-ation (squeezed) states. All these show \emph{monotonic} and \emph{unlimited} increases of
values of $\Xi$ with increases of their nonclassical character. Additionally, \cat states allow us
to show, that $\Xi$ drops with drops of coherences $|\state|$ and rises quadratically with \ps
distances~$y$ in~$\varrho(x-y,x+y)$.

We also prove that increasing the mixedness of a quantum state reduces $\Xi$ in the sense that it is
a convex function of the mixing probabilities, see Eq.~(\ref{eq:MoreImpurity_LessXi}) below.

\section{Wigner's Phase Space Distribution\label{sec:Liouville_W_dist}}

$W(x,p)$, for a one-dimensional continuous system's quantum state~$\state$, is given by the Fourier
transform with respect to the distances~$y$ between its off-diagonal coherences ${\varrho(x-y,x+y)}$
$ = \langle x-y | \hat \varrho |x+y
\rangle$~\mbox{\cite{Wigner_PR32,Daubechies__JMP83,Hillery_PR84}}
\begin{eqnarray}\label{eq:_WignerDistr}
  W(x,p) = \frac{1}{\pi} \int_{-\infty}^{\infty} dy
  \; \varrho(x-y,x+y) \; {\rm e}^{{2 {\rm i}} p y}.  
  \quad 
\end{eqnarray}
By construction (Sect.~\ref{subsec:Liouville_W_dist}), $W$ is normalized and nonlocal
(through~$y$). Whereas~$\varrho$ tends to be complex-valued, $W$ is always real-valued but,
generi\-cally, $W$ has nega\-tive values in some regions of
\ps~\cite{Wigner_PR32,Hillery_PR84}. Since~$W$ and~$ \varrho$ are Fourier transforms~(\ref{eq:_WignerDistr}) of
each other, they are unitarily equivalent to each other.

Because $W$ and $\state$ are unitarily equivalent and $Q$ is just a smeared version
of~$W$~\cite{Cahill_PR69b,Schleich_01}, see \ref{subsec:S-parametrized_dist}, we use $\Xi[W]$ and
$\Xi[\state]$ interchangeably.

Unlike classical Liouville \ps probability densities, $W$ is not a probability
distribution~\cite{Mueckenheim__PR86}, this is most clearly seen from the fact that
frequently~$W(x,p) < 0$; if this happens the state is nonclassical.  But there are states which are
nonclassical and yet~${W(x,p) \geq 0}$~\cite{Wuensche_JOBQSO04}, that is why we use Bohmann and
Agudelo's certification function~$\xi$. Since $W$ and~$Q$ exist for all states and are always well
behaved, so do $\xi[W]$ and $\Xi[W]$.

\subsection*{Other Phase Space Distributions\label{subsec:S-parametrized_dist}}

There are families of \ps distributions such as the Glauber-Sudarshan distribution $P={\cal P}_1$, the
Wigner distribution~$W={\cal P}_0$ and the Husimi-Q distribution
$ Q = {\cal P}_{-1}$~\cite{Cahill_PR69b}. It is known that they are connected through a smoothing
convolution-identity~\cite{Bohmann_PRL20,Cahill_PR69b}.  Here, since we work in (quantum-mechanical)
\ps, we follow the standard convention for distributions in cartesian \ps, see
Ref.~\cite{Leonhardt_PQE95} and Sect.~\ref{subsec:S-parametrized_dist}.

\begin{widetext}
\section{Growth of Quantumness with Growth of Mutual Coherence\label{sec:_Xi_grows_coherence}}

The nonclassical features displayed by quantum \ps distributions are due to coherences embodied by
the two-point correlation-function $\varrho(x-y,x+y)$ used for the construction
of~$W$~(\ref{eq:_WignerDistr}). The coherences give rise to interference patterns in
\ps~\cite{Schleich_01} whose fringes,~see Fig.~\ref{Fig:details_cat}, are most transparently studied
for states with only two narrow regions of support, so called \cat states.
A special case of such pure two-component \cat states are those formed from an unbalanced
superposition of two Glauber coherent states~\cite{Schleich_01,Vogel_Welsch__QOBook06}, centered on
\ps
locations $(x,p)=(0,{x_0})$ and $(0,-{x_0})$, respectively, given by
\begin{align} \label{eq:App_Psi_cat} \Psi_{\text{\cat}}(x; {x_0},w,\theta) & = {\rm
      e}^{-\frac{x^2}{2}} \left(\cos (w) {\rm e}^{ {\rm i} x {{x_0}}}+{\rm e}^{{\rm i}\theta }\sin
      (w){\rm e}^{-{\rm i} x {{x_0}}} \right) / \sqrt{\sqrt{\pi}({\rm e}^{-{{x_0}}^2} \sin (2 w)
      \cos (\theta )+1)} \, , \\
  \label{eq:App_W_cat_main_text}
  \text{for which }\quad  W_{x_0}(x,p;{x_0},w,\theta) & = \frac{{\rm e}^{-p^2-x^2}
    \left({\rm e}^{+{{x_0}}^2} \sin (2 w) \cos (2 {{x_0}} x-\theta )+{\rm e}^{2 p {{x_0}}} \cos
      ^2(w) + {\rm e}^{-2 p {{x_0}}} \sin ^2(w)\right)}{\pi \left({\rm e}^{+{{x_0}}^2}+\sin (2 w)
      \cos (\theta )\right)} \; .
\end{align}
The relative phase angel, $\theta$, ranges over $[-\pi,\dots,\pi ]$. \mbox{$\theta = 0$}, creates an even
\cat, $\theta = \pi$ an odd \cat state~\cite{Kenfack_JOB04}.

The weighting angle, $w$, ranges over $[0,\dots, \frac{\pi}{2} ]$, $w$ parameterizes the balance
between the components and thus their mutual coherence amplitude~$\sin(2w) = 2 \sin(w) \cos(w)$ as
the coefficient of the \ps interference term~$\sin(2w) \times \cos (2 {x_0} x - \theta)$
in~(\ref{eq:App_W_cat_main_text}), which has full contrast when balanced, that is, for
$w = \frac{\pi}{4}$ the cohe\-rence amplitude is~$\sin(2 \frac{\pi}{4}) = 1$.

It can be shown analytically, for \mbox{amplitudes ${x_0 \gtrsim 3}$} and all relative
angles~$\theta$, that moving the mixing angle away from the balancing point~$w \neq \frac{\pi}{4}$
reduces the magnitudes of the negative values of $W$ and $\xi$, leading to a \emph{monotonic}
reduction in~$\Xi$ with decreasing coherence, parameterized by increasing distance of~$w$ from the balancing point $w=\pi/4$.
\end{widetext}

\section{Quadratic Scaling of $\Xi$ with $x_0$\label{sec:_Xi_grows_quadratically}}

To study the quadratic dependence on distance in \ps we next focus on balanced \cat states,
with~$w = \frac{\pi}{4}$, their Wigner distribution~(\ref{eq:App_W_cat_main_text}) is
\begin{equation} \label{eq:pure_W_cat}
\!\!\!  W_{x_0} (\theta) = \frac{{\rm e}^{-x^2 -p^2} \! \left({\rm e}^{{x_0}^2}\! \cos (2 {x_0}
               {{x}} - \theta )+ \cosh(2 {{p}} {x_0}) \right)}{
               \pi  \left({\rm e}^{+{x_0}^2} + \cos (\theta )\right)}.\!\!                           
\end{equation}

\begin{figure}[b] \centering
  \includegraphics[width=2.8cm,height=2.3cm]{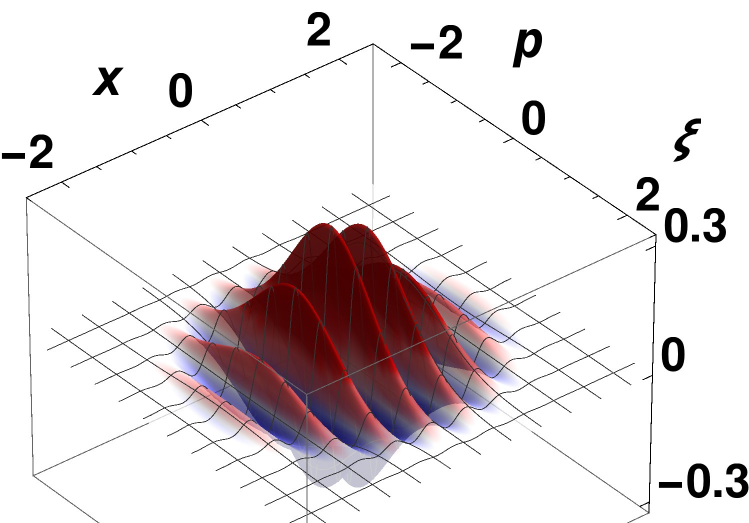}
  \includegraphics[width=2.8cm,height=2.3cm]{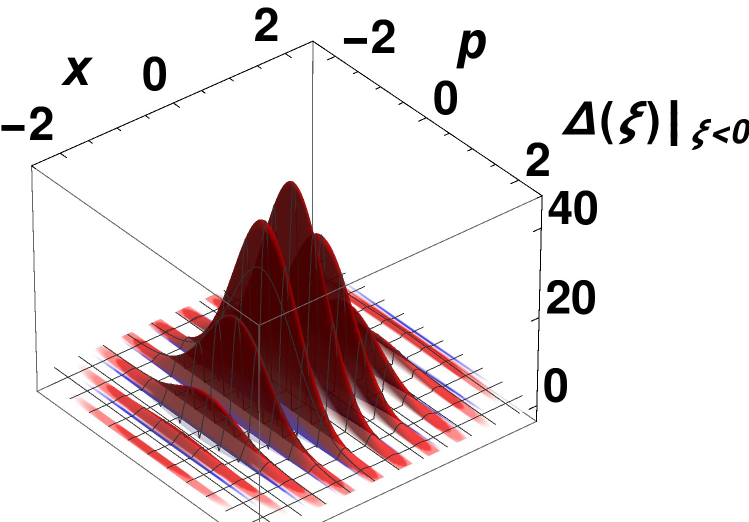}  
  \includegraphics[width=2.8cm,height=2.3cm]{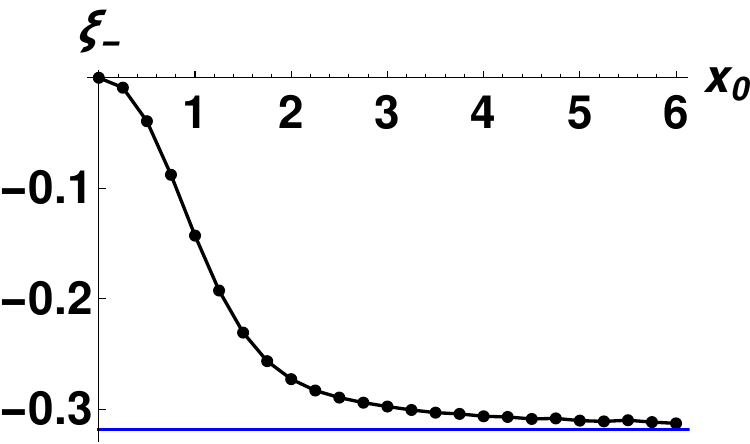}  
  \caption{Left panel: $\xi(x,p)$, around the origin, for an odd \cat
    state~$W_{x_0=6}(\theta = \pi)$.  Middle panel: local quantumness $\VEC{\Delta}\xi(x,p)|_{\xi<0}$
    for~$W_{6}(\pi)$.  Right panel: trend line for minimum value $\xi_-[W_{x_0}(\pi)]$ for odd
    \cat states with \cat size~$2\, x_0$.
    \label{Fig:details_cat}}
\end{figure}

An increase in the ``effective size’’ of the \cat state due to an increase in the
displacements~${x_0}$ leads to an increase in the spatial frequency (how dense the fringes are) in
the interference term $\cos (2 {x_0} {{x}} - \theta )$ of~(\ref{eq:pure_W_cat}).
Dense fringes easily get washed away by the smoothing in the mapping
from $W_{x_0}$ to $Q_{x_0}$~\cite{Schleich_01}. This is why above~$x_0 \approx 6$ the $\xi_-$-plot
in Fig.~\ref{Fig:details_cat} saturates,  reflecting the
fact that near the origin, $\xi$'s $Q^2$-term~(\ref{eq:xi_Certification})
drops to negligible values.
Addi\-tio\-nally, for coherent states ${\xi \geq 0}$~\cite{Ole_25_TowardsQuantumness}. Hence,
for~$W_{x_0}$ the certification function $\xi(x,p)$ is non-negative everywhere in \ps except for the
interference region around the origin. We can therefore exclusively concentrate on the `nonclassical
basins' ($\xi < 0$; see Fig.~\ref{Fig:details_cat}).

\begin{figure}[t] \centering
  \includegraphics[width=2.8cm,height=2.8cm]{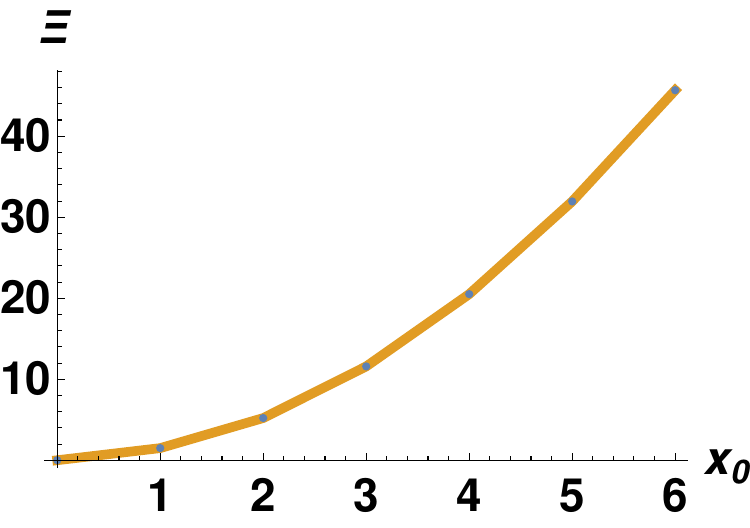}
  \includegraphics[width=2.8cm,height=2.8cm]{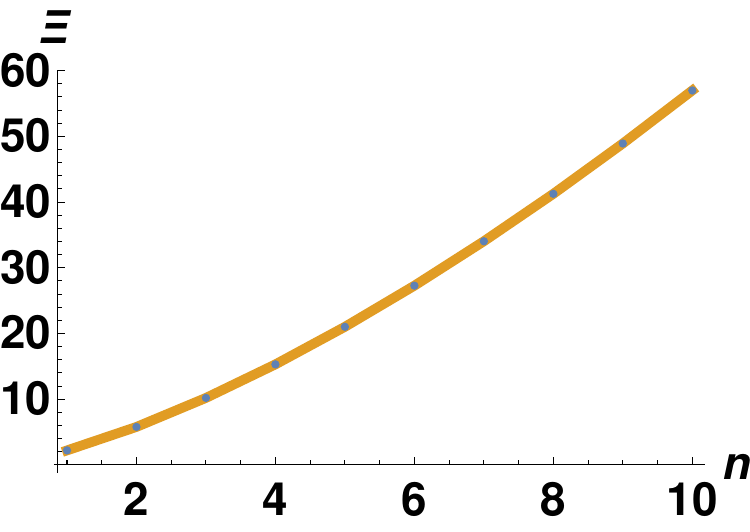}
  \includegraphics[width=2.8cm,height=2.8cm]{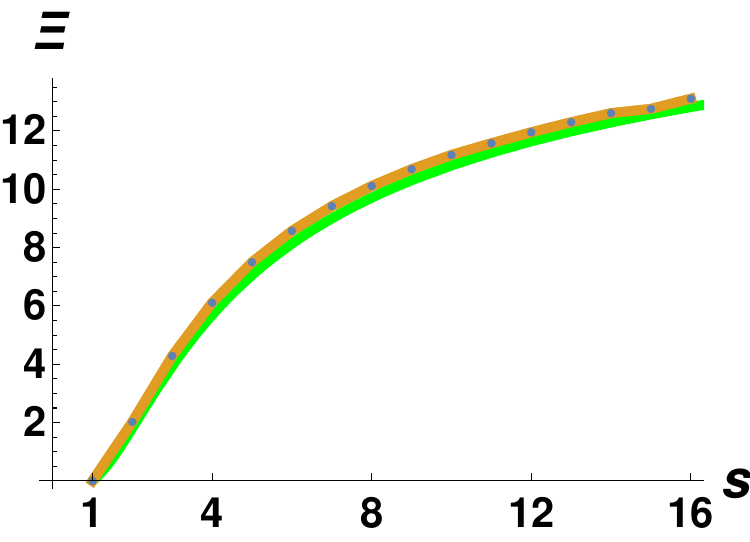}
  \caption{Left to right: trend lines of~$\Xi$ for positive
    \cat,~$W_{x_0}$ (compare Fig.~\ref{fig:Xi_I_pure_cats}), Fock states,~$W_{|n\rangle}$,
    and pure squeezed states,~$W_{\squ}$.
    The plots    confirm the monotonic growth with increasing quantumness of the states. The thick green curve
    for $\Xi[W_s]$ is the analytical expression discussed in~\ref{sec:PureSqueezedState}.
    \label{Fig:Trends_Xi}}
\end{figure}

Fig.~\ref{Fig:Trends_Xi} shows that the quantumness measure for balan\-ced \cat states,
$\Xi[W_{x_0}]$, grows monotonically with its effective size $2 {x_0}$. A good approximation is
\begin{flalign}
  \Xi[&W_{x_0}(\theta)]  \approx \iint \! dp \, dx \; \big.\VEC{\Delta} W (x,p)\Big|_{W<0} \approx \notag \\
  & \frac{1}{\pi}\iint \! dp \, dx \; {{\rm e}^{-x^2 -p^2}} \cos(2x_0p)|_{\cos <0} =\frac{4}{\pi}
  x_0^2\; ,
  \label{eq:Xi_Measure_pure_CatStateApproximation} 
\end{flalign}
which applies once $x_0$ is large enough (see Fig.~\ref{fig:Xi_I_pure_cats}) to avoid \ps
overlaps between Glauber states and their \ps interference pattern depicted in
Fig.~\ref{Fig:details_cat}.

This quadratic scaling, $\Xi[W_{x_0}] \sim {\cal O}(x_0^2)$, arises because the \emph{local
  quantumness} term in~(\ref{eq:Xi_Measure}) is designed to behave `\mbox{kinetic}
energy'-like~\cite{Lee_Jeong__PRL11}.  With increasing amplitude~$x_0$, the {\cat}s interference
fringes narrow inversely with~$x_0$, yielding quadratic growth of~$\Xi$ with~$x_0$, see
Fig.~\ref{Fig:Trends_Xi} and~\ref{fig:Xi_I_pure_cats}.

\section{Widely Coherently Spread-Out States: Unboundedness of $\Xi$\label{sec:_Xi_grows_unbound}}

\begin{figure}[h] \centering
  \includegraphics[width=2.2cm,height=2.2cm]{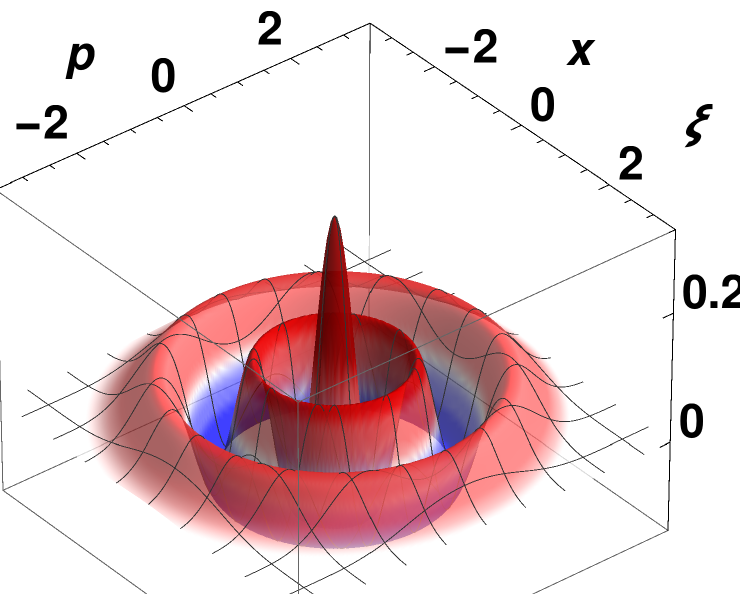}
  \includegraphics[width=2.8cm,height=2.2cm]{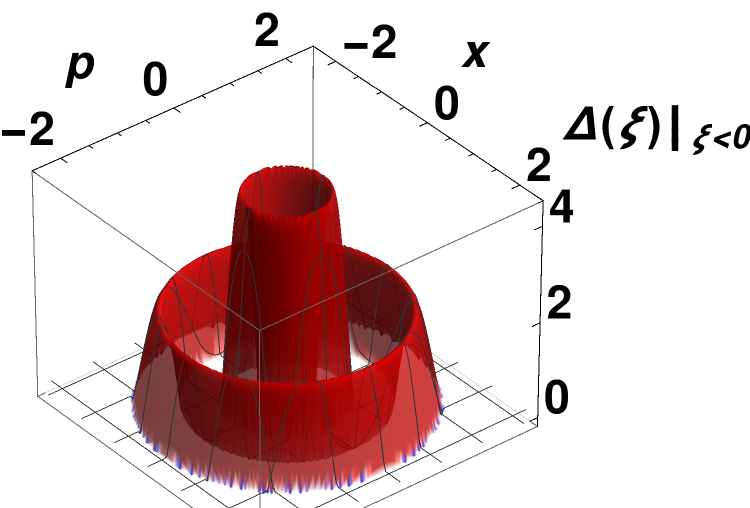}  
  \includegraphics[width=3.4cm,height=2.2cm]{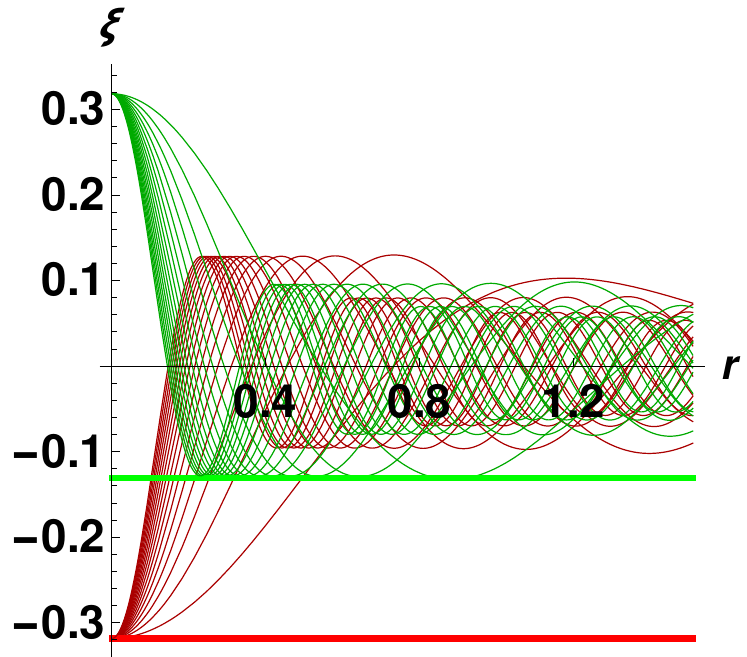}
  \caption{From left to right: Values of $\xi(x,p)$ for fourth excited Fock state~$W_{|4\rangle}$. Values
    of local quantumness $\VEC{\Delta}\xi(x,p)|_{\xi<0}$ for~$W_{|4\rangle}$. Values of
    $\xi(r)=\xi(\sqrt{x^2+p^2})$, up to quantum number $n=34$, for odd Fock states (red graphs) and
    even Fock states (green graphs), displaying respective minimum values \emph{oscillating} between
    $\xi_-^{\rm odd}=-1/\pi$ and $\xi_-^{\rm even}\approx -0.131$.
    \label{Fig:details_fock}}
\end{figure}

To study the \emph{monotonic} and \emph{unbounded} growth of $\Xi$ for widely coherently spread-out
states, we next consider harmonic oscillator energy eigenstates, also known as `Fock' states, or
`number states'~\cite{Schleich_01,Vogel_Welsch__QOBook06}, Fig.~\ref{Fig:details_fock} illustrates
the behaviour of~$\xi$ for them.  With every increase of their \mbox{energy} \mbox{quantum} number,
$n$, Fock states,~$| n \rangle$, extend further across \ps.

\begin{figure}[h]
  \hspace{-0.2cm}
  \begin{minipage}[b]{\columnwidth}
    \; \includegraphics[width=0.95
    \columnwidth,height=2.05cm]{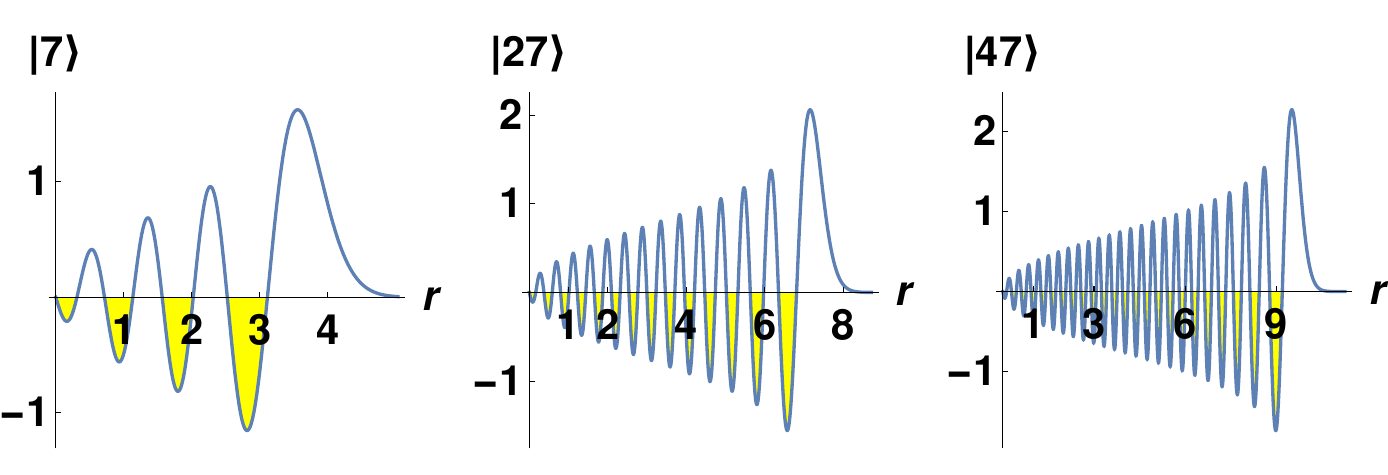}
    \caption{Normalized radial density $2\pi\, r\, W_{|n\rangle} (r)$ for Fock states $|7\rangle$,
      $|27\rangle$ and  $|47\rangle$. Because of the rotational \ps symmetry of Fock states
      $ \iint_{-\infty}^{\infty} dx dp \, W_{|n\rangle}
      = \int_{0}^{\infty} dr \, 2\pi r\, W_{|n\rangle} = 1$.
      \label{fig:WignerToInfinityPlot}}
\end{minipage}
\end{figure}

States with unbounded magnitude, \mbox{$\iint |W|\; dx dp \rightarrow \infty$},
exist~\cite{Wlodarz__IJTP03}. Fock states with `infinite energy'
belong to this class, see Fig.~\ref{fig:WignerToInfinityPlot}, and therefore, to remain normalized,
their negative weight in \ps has to keep growing with~$n$:
$\lim_{n \rightarrow \infty} \iint W_{|n\rangle}|_{W<0} \; dx dp = -\infty$.
Consequently, their quantumness~$\Xi$ keeps growing with increasing energy.

This growth of quantumness~$\Xi[W_{|n\rangle}]$ with $n$ is depicted in Fig.~\ref{Fig:Trends_Xi}. We
could not find an analytical expression to describe it in closed form.

More general, coherently widely spread-out states commonly occur in weakly confined systems. If
pure, they show far-reaching interference across \ps~\cite{Ole_JPAMT23}. The further they are spread
out, the larger the surface of areas where~$W$ is negative~\cite{Ole_JPAMT23}. And the area where
$\xi$ is negative grows with it (since $Q$ is positive and normalized~\cite{Schleich_01} and thus
dilutes, unlike~$W$). Therefore, just like for Fock states, the further spread out general pure
states, such as those in Ref.~\cite{Ole_JPAMT23}, are, the more their quantumness tends to increase,
\emph{without bound}.

\section{Squeezed States: Unbounded $\Xi$\label{sec:_Xi_grows_squeezedStates}}

To study the \emph{monotonic} and \emph{unbounded} growth of $\Xi$ for coherently spread-out states
which are at the same time narrowly squeezed, we consider squeezed states~\cite{Lvovsky_16squeezed}.
These states arguably show quantumness in a somewhat subtle way, they form positive Wigner
distributions, of Gaussian form~\cite{Hudson_RMP74} but their quadrature fluctuations can be below
that of the vacuum state-threshold~\cite{Ole_25_TowardsQuantumness}.

\begin{figure}[t] \centering
  \includegraphics[width=2.5cm,height=2.2cm]{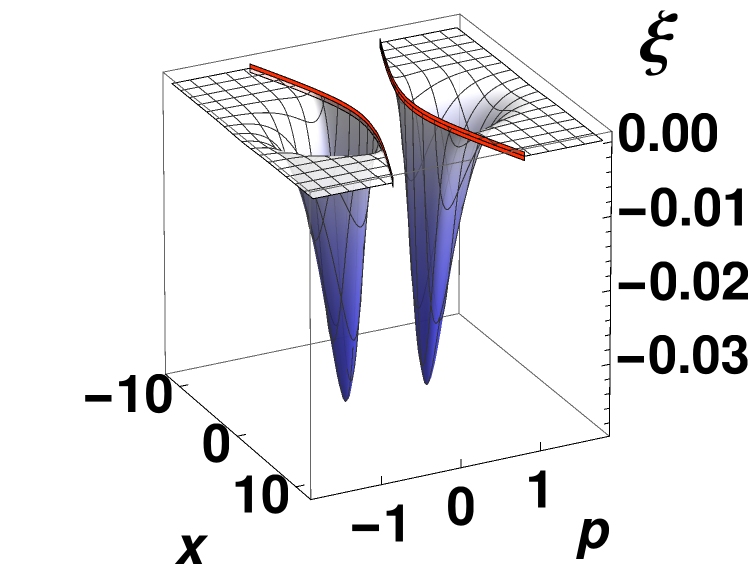}
  \includegraphics[width=2.5cm,height=2.2cm]{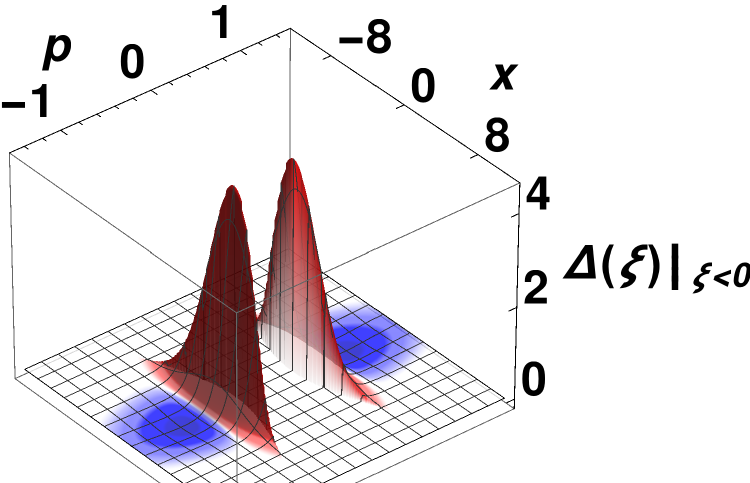}  
  \includegraphics[width=3.4cm,height=2.2cm]{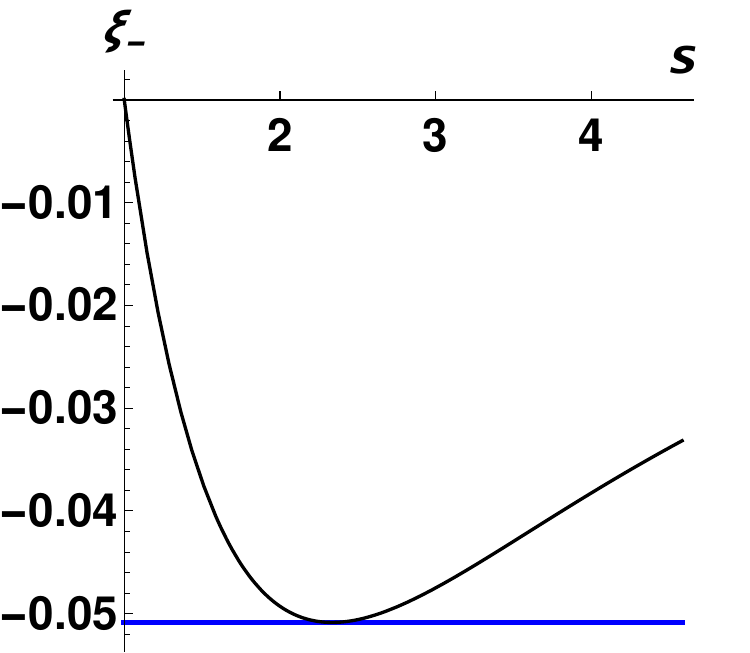}
  \caption{ Left panel: values of $\xi(x,p)|_{\xi<0}$ for a pure squeezed
    state~$W_{\squ=4}^{[\mu=1]}$~(\ref{eq:W_impure_squeezed}), the zero contour is highlighted.
    Middle panel: values of local quantumness $\VEC{\Delta}\xi(x,p)|_{\xi<0}$ for the same
    state.  Right panel: trend line for minimum value
    $\xi_-[W_{\squ}^{[\mu=1]}]$ for pure squeezed states~(\ref{eq:W_impure_squeezed}).
    \label{Fig:details_squeezed}}
\end{figure}

When this below-threshold squeezing of the fluctuations happens, the states are
nonclassical~\cite{Wuensche_JOBQSO04}. For all Gaussian states, this is always certified by
$\xi$~\cite{Ole_25_TowardsQuantumness}, see Fig.~\ref{Fig:details_squeezed}, and $\Xi$ quantifies
it, see Fig.~\ref{Fig:Trends_Xi}.

The Wigner distribution of \emph{impure} squeezed states (centered on the origin and rotated to
align with the \mbox{$x$-axis}) has the Gaussian form
\begin{eqnarray}
  \label{eq:W_impure_squeezed} W_{\squ}^{[\mu]} = \frac{1}{\sqrt{\mu} \pi}
  \exp(- \frac{x^2}{\mu V_s} - V_s \; p^2) \; ,
\end{eqnarray} where the squeezing parameter, $\squ = \sqrt{V_s}$, is rescaled in terms of the
variance of the vacuum state $W_{\squ = 1}^{[\mu=1]}$ and values $\mu >1$ parameterize the impurity
of the state, for further details see~\ref{sec:ImpureSqueezedState}.

Figure~\ref{Fig:Trends_Xi} shows that the quantumness measure~$\Xi[W_{\squ}]$ for pure squeezed
state~$W_{\squ}$ grows monotonically, despite the fact that the magnitude of~$\xi_-[W_{\squ}]$
decreases for large values of squeezing~$\squ$, see Fig.~\ref{Fig:details_squeezed}.  Specifically,
the growth of~$\Xi[W_{\squ}]$ for large values $\squ$ is \emph{unbounded}. For pure states, in
leading order, it is
$ \Xi[W^{[\mu=1]}_\squ] \approx 16 \left({\frac{1}{\pi} \log (\frac{\squ }{2})}
\right)^{\frac{1}{2}}$. For further details see
Fig.~\ref{Fig:Trends_Xi_squeezed_numerical_approximate} and Appendix~\ref{sec:PureSqueezedState}.

\section{Discriminating Mixed States \label{sec:_mixed_states}}

\begin{figure}[t] \centering
  \includegraphics[width=4.05cm,height=2.5cm]{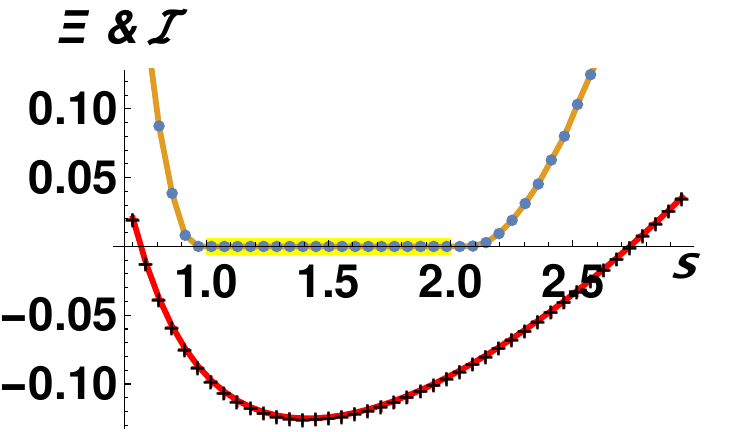}
  \includegraphics[width=4.05cm,height=2.5cm]{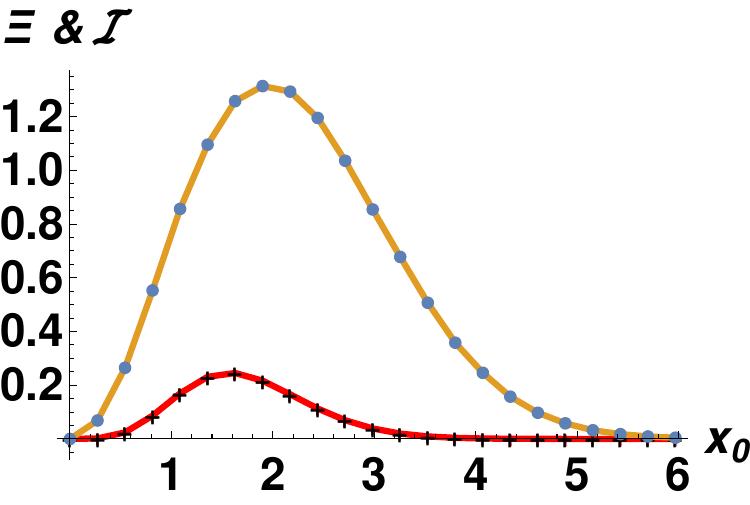}
  \setlength{\fboxsep}{0pt}
  \begin{picture}(0,0)
\put(-65,40){\fcolorbox{brown}{white}{\includegraphics[height=1.1cm]{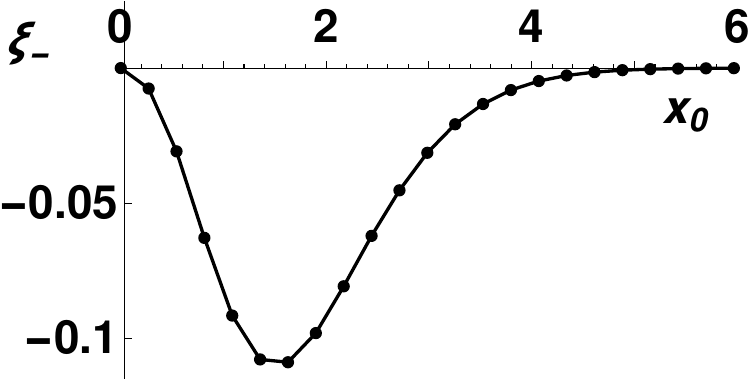}}}
\end{picture}
\caption{Values of $\Xi$ (light brown curves with dots), Eq.~(\ref{eq:Xi_Measure}), and
  ${\cal I}$~\cite{Lee_Jeong__PRL11} (dark red curves with crosses). Left panel: \emph{impure}
  squeezed states~$ W_{\squ}^{[\mu=4]}$~(\ref{eq:W_impure_squeezed}), with an impurity factor
  of~$\mu = 4$ are known to be classical~\cite{Wuensche_JOBQSO04} in the yellow-highlighted region
  $\squ = 1,\ldots, 2$.  Whereas $\Xi$ behaves suitably, ${\cal I}$~does not. Right panel:
  \emph{impure} Glauber-\cat states [$W_{x_0}(\theta=\tfrac \pi 2 )$ with \mbox{vacuum} mixed in via
  a 25\% reflec\-tivity beam splitter, other mode traced out].  For values of $x_0$, above
  $x_0\approx 6$, the interference fringes at the origin are washed out rendering these states
  classical. Whereas $\Xi$ behaves suitably, as certified by~$\xi_-$ (see inset), ${\cal I}$ is
  insensitive to the presence of nonclassical behaviour for $x_0$-values below 0.25 and above 3.5,
  roughly.
  \label{Fig:Xi_vs_LeeMeasure}}
\end{figure}

For mixed states we observe that $\Xi$ drops with increa\-sing impurity, see
Figs.~\ref{Fig:Xi_vs_LeeMeasure} and~\ref{fig:Xi_Lee_Thermal}.

This is due to that fact that increasing impurity, by incoherently adding the state $W_1$ with
probability ${0 < {w} <1}$ to state $W_2$, reduces nega\-tive parts of~$\xi$, as the
positive term $ 4\pi {w} (1-{w}) (Q[W_1] - Q[W_2])^2$ gets added:
\begin{flalign}
  \!\!\!  \xi[{w} W_1 + (1-{w}) W_2] = & \; {w} \, \xi[W_1] + (1-{w}) \; \xi[W_2]  \notag \\
  + \; 4 \pi & {w} (1-{w}) (Q[W_1] - Q[W_2])^2 ,
  \label{eq:MoreImpurity_Lessxi} 
\end{flalign}
rendering~$\Xi$ of~(\ref{eq:Xi_Measure}) convex in ${w}$: 
\begin{flalign}
\!\!\!  \Xi[{w} W_1 + (1-{w}) W_2] < & \; {w} \, \Xi[W_1] + (1-{w}) \; \Xi[W_2] \, .
  \label{eq:MoreImpurity_LessXi} 
\end{flalign}
We have found only two other measures, see~\ref{subsec:OtherMeasures}, that are \emph{\cti},
\emph{monotonic} and \emph{unbounded}:~${\cal I}$~\cite{Lee_Jeong__PRL11}, of
Eq.~(\ref{eq:LeeJeong_Measure}) below, and an entropic measure~\cite{Zhang_Luo__EPJP21}. But neither
is as \emph{sensitive} for mixed states as~$\Xi$, see
Figs.~\ref{Fig:Xi_vs_LeeMeasure},~\ref{fig:Xi_Lee_Thermal}
and~\ref{Fig:not_discriminating_Zhang_Luo}.

${\cal I}$ of Ref.~\cite{Lee_Jeong__PRL11} has the fixed form
\begin{flalign}
  {\cal I}[W] = -\, \pi \iint_{-\infty}^{\infty} dx \, dp \; \; W \left( \frac{{\VEC
        \Delta} W}{2}+W \right) .
  \label{eq:LeeJeong_Measure} 
\end{flalign}

\begin{figure}[h!]
  \begin{minipage}[b]{0.99\columnwidth}
    \includegraphics[width=4.20cm,height=4.075cm]{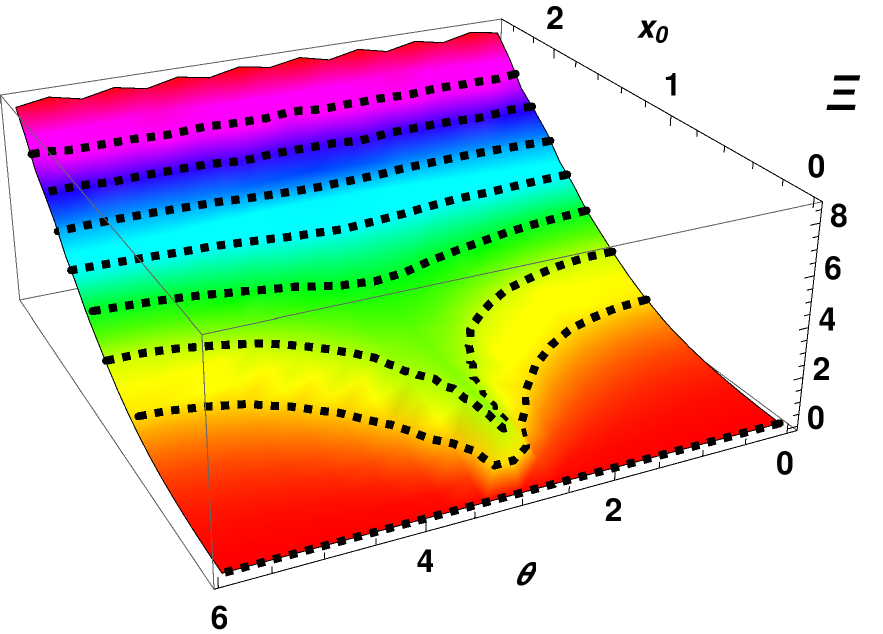}
    \includegraphics[width=4.20cm,height=4.075cm]{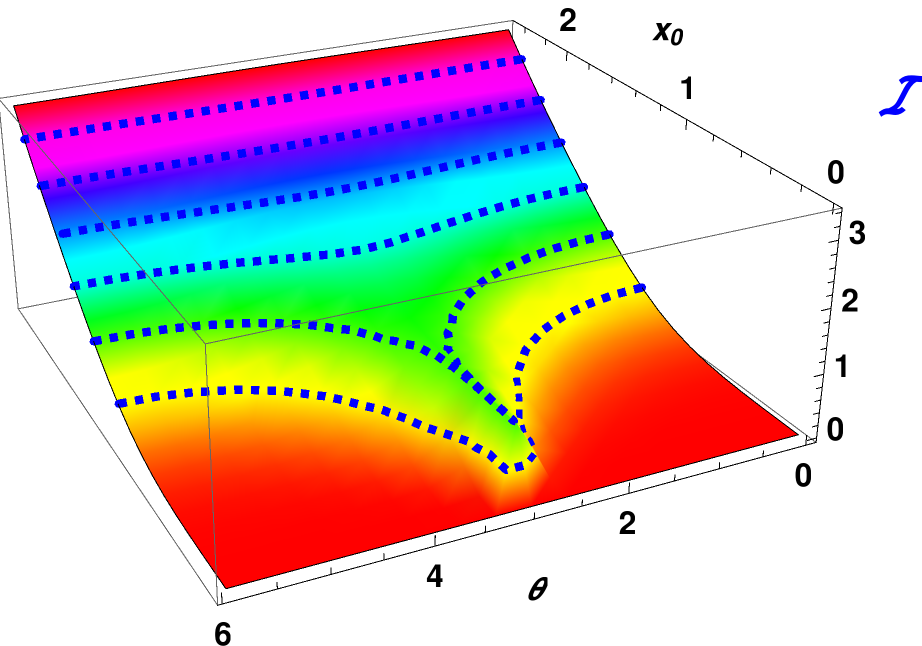}
    \caption{Quantum measures $\Xi$ of Eq.~(\ref{eq:Xi_Measure}) and $\cal I$ of
      Ref.~\cite{Lee_Jeong__PRL11} (also see Eq.~(\ref{eq:LeeJeong_Measure})), for pure cat
      states,~$ W_{x_0} (\theta)$,~(\ref{eq:pure_W_cat}) with peak distances $2 x_0$ and relative
      phase, $\theta$, between peaks.  `Negative cats'~(${\theta = \pi}$) are slightly more
      non-classical than `positive cats' ($\theta = 0$). Both, $\Xi$ and $\cal I$, show smooth
      monotonic growth in~$x_0$. Above $x_0 \approx 1.5$,~$\Xi \sim {\cal O}(x_0^2)$ for all~$\theta$,
      compare Eq.~(\ref{eq:Xi_Measure_pure_CatStateApproximation}).
      \label{fig:Xi_I_pure_cats}}
  \end{minipage}
\end{figure}

In our minds, $\cal I$ is the best competitor of $\Xi$, but for mixed states measure~${\cal I}$ is
less sensitive than~$\Xi$ when discri\-mi\-na\-ting between classical and nonclassical states, see
Figs.~\ref{Fig:Xi_vs_LeeMeasure} and~\ref{fig:Xi_Lee_Thermal}. For large quantumness, $\Xi$ and
$\cal I$ behave qualita\-tively similarly to each other, see
Figs.~\ref{fig:Xi_I_pure_cats},~\ref{fig:XiMeasure_I_sq_asq} and~\ref{fig:Xi_I_landscape_dB}.

\section{Application to Experimental Data \label{sec:_apply_experimentatl_data}}

In our experiments~\cite{Hsieh__PRL22} with large squeezing we can encounter large
impurities. Existing measures are not \emph{guaran\-teed} to be sensitive enough to
{discriminate} between classical and nonclassical behaviour~\ref{sec:_mixed_states} but
measure~$\Xi$~(\ref{eq:Xi_Measure}) works faithfully for gaussian states, see
Ref.~\cite{Ole_25_TowardsQuantumness}.

\begin{figure}[t]
  \hspace{-0.2cm}
  \begin{minipage}[b]{\columnwidth}

    \includegraphics[width=0.7
    \columnwidth,height=3.085cm]{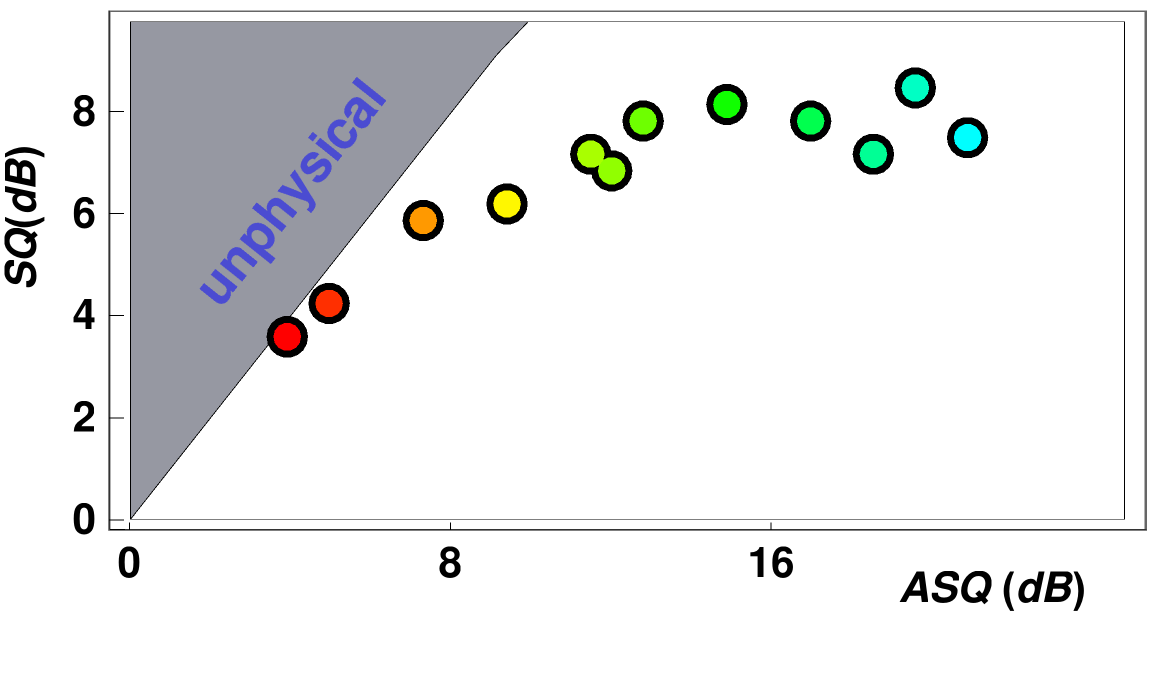}
    \caption{For impure squeezed states $W(x,p,\mu V_\squ,1/V_\squ) $ (\ref{eq:W_impure_squeezed}):
     the top panel shows experimental data from Fig.~1 of Ref.~\cite{Hsieh__PRL22}, the gray
     area is unphysical since it would imply that $V_x V_p < 1$. 
    \label{fig:Xi_I_sq_asq}}
\end{minipage}
\end{figure}

\begin{figure}[t]
  \hspace{-0.2cm}
  \begin{minipage}[b]{\columnwidth}
    \includegraphics[width=0.7 \columnwidth,height=3.085cm]{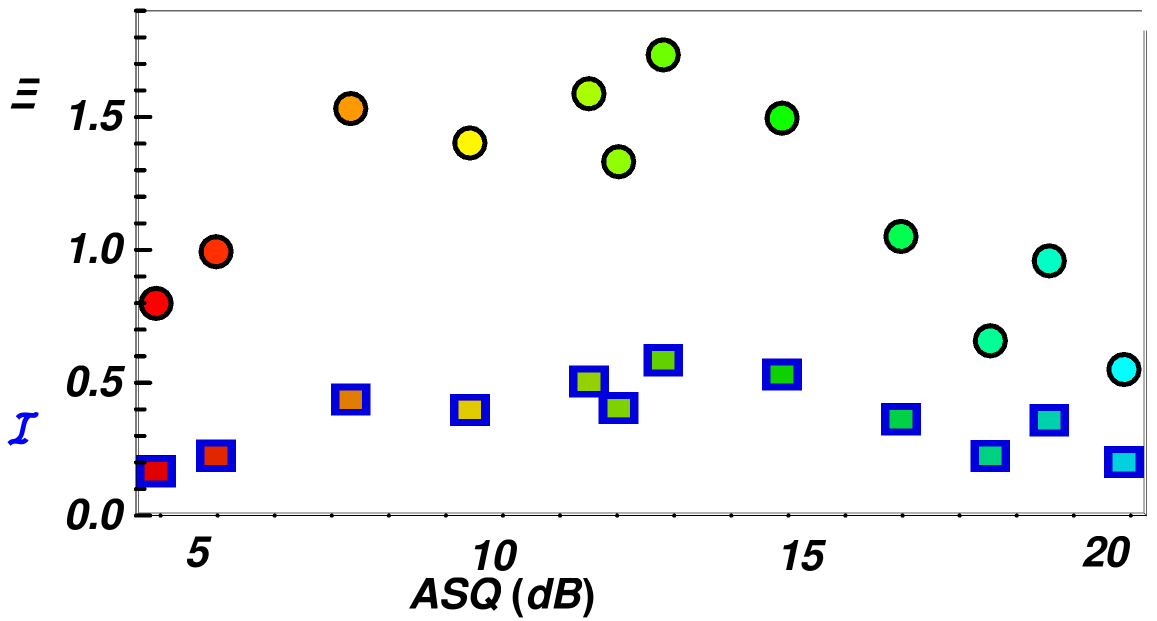}
    \caption{ Quantumness values (black circles for measure $\Xi$~(\ref{eq:Xi_Measure}) and blue
      squares for measure ${\cal I}$ of Eq.~(\ref{eq:LeeJeong_Measure})) for the impure squeezed
      states of Ref.~\cite{Hsieh__PRL22} displayed in Fig.~\ref{fig:Xi_I_sq_asq}, above (for further
      details see~\ref{sec:ImpureSqueezedState}).
      \label{fig:XiMeasure_I_sq_asq}}
\end{minipage}
\end{figure}

Fig.~\ref{fig:Xi_I_sq_asq} shows states with various degrees of squeezing and
impurity~\cite{Hsieh__PRL22}. Before we found the certification functional~$\xi$ and based the
measure~$\Xi$ on it, it was unclear how to reliably quantify the quality of our experimentally
generated squeezed states, since no guaranteed measure of their quantumness was available. Here, we
can now close this gap, a quantifi\-cation is given in Fig.~\ref{fig:XiMeasure_I_sq_asq}.  This
enables us to reliably~\cite{Ole_25_TowardsQuantumness} compare the quantumness of our
experimentally generated states.


\section{Other Quantumness Measures \label{sec:_otherMeasures_etc}}

In Appendix~\ref{subsec:OtherMeasures} we consider several other quantumness measures, none of
these~\cite{Hillery__PRA87,Lee__PRA91,Lee__PRA92,Kenfack_JOB04,Bjoerk__JOBQSO04,Luetkenhaus__PRA95,Manfredi__PRE00,Dodonov_Wuensche__JMO00,Shimizu__PRL02,Marian__PRL02,Duer_Cirac_PRL02,Dodonov_Reno__PLA03,Malbouisson__PS03,Asboth__PRL05,Cavalcanti_Reid__PRL06,Cavalcanti_Reid__PRA08,Boca__PRA09,Lee_Jeong__PRL11,Froewis_Duerr__NJP12,Nimmrichter_Hornberger_PRL13,Yadin_Vedral__PRA16,Tan_Jeong__PRL17,Kwon_Jeong_NJP17,Sekatski__NJP18,Tan_Jeong__PRL20,Zhang_Luo__EPJP21,vanHerstraeten_Cerf__PRA21,Naseri__PRA21}
are simul\-taneously \emph{\sens, \cti, monotonic}, and \emph{unbounded}.

We note, there has at times been some confusion due to a lack of
distinction~\cite{Luetkenhaus__PRA95,Asboth__PRL05} {between the quantification of} the size of
quantumness --using a
\mbox{measure~\cite{Hillery__PRA87,Lee__PRA91,Lee__PRA92,Kenfack_JOB04,Bjoerk__JOBQSO04,Luetkenhaus__PRA95,Manfredi__PRE00,Dodonov_Wuensche__JMO00,Shimizu__PRL02,Marian__PRL02,Duer_Cirac_PRL02,Dodonov_Reno__PLA03,Malbouisson__PS03,Asboth__PRL05,Cavalcanti_Reid__PRL06,Cavalcanti_Reid__PRA08,Boca__PRA09,Lee_Jeong__PRL11,Froewis_Duerr__NJP12,Nimmrichter_Hornberger_PRL13,Yadin_Vedral__PRA16,Tan_Jeong__PRL17,Kwon_Jeong_NJP17,Sekatski__NJP18,Tan_Jeong__PRL20,vanHerstraeten_Cerf__PRA21,Naseri__PRA21}--}
on the one hand and certification of the presence of non\-classicality, on the other hand,
see~\cite{Vogel__PRL00,Diosi__Vogel__PRL00_Comment,Vogel__PRL00_Reply,Shchukin_Richter_Vogel__PRA05,Miranowicz_PRA10,Bohmann_Q20,Park_PRR21}
and the many references therein.

\section{Can one number quantify  quantumness? \label{subsec:_one_number}}

It was speculated that ``\emph{the Hilbert space containing all pure states is huge}'' and
``\emph{so one should not expect to characterize the non-classical features of a quantum state just
  by a single scalar quantity}''~\cite{Kenfack_JOB04}. We disagree, our measure successfully yields
just that, a single scalar quantity, and so do other measures such as
$\cal I$~(\ref{eq:LeeJeong_Measure}) of Ref.~\cite{Lee_Jeong__PRL11} or
decoherence~\cite{Nimmrichter_Hornberger_PRL13} or entropic measure~\cite{Zhang_Luo__EPJP21}.

Indeed, Fig.~\ref{fig:Xi_I_pure_cats} demonstrates that $\Xi[W_{x_0}]$, for all phase
angles~$\theta$, grows \emph{monotonically} with~$x_0$ and thus smoothly, and keeps growing without
bound.  Contrast this, e.g., with the non-monotonic behaviour of the measure constructed in
Ref.~\cite{Kenfack_JOB04}.  We believe that \emph{monotonic} growth facilitates smoothness and this
way makes a measure useful for comparison across all types of states.

\section{Quadratic scaling with \ps distance might enforce uniqueness \label{subsec:_unique}}

$\Xi = 0$ for all pure coherent states~\cite{Bohmann_PRL20,Ole_25_TowardsQuantumness}. In a manner
of speaking this fixes the \emph{universal} `floor' of $\Xi$ at zero. Having established that $\Xi$
grows when the quantumness of the state increases and does so without bound, we have, in a manner of
speaking, also fixed the universal `ceiling' of $\Xi$, at `infinity'. To our knowledge, $\Xi$ is the
only measure which is thus simultaneously \emph{\sens, \cti}, \emph{monotonic} and \emph{unbounded}.

Since the `ceiling' of a quantumness measure should be \emph{unbounded} or infinite, approaches
different from this work, possibly using different scaling when the measure grows, might be
feasible.  But, Ref.~\cite{Nimmrichter_Hornberger_PRL13} gives \mbox{general} arguments to support
growth quadratic with the effec\-tive size of a quantum state. Similar arguments based on decoherence
mechanisms in Ref.~\cite{Lee_Jeong__PRL11} also yield a measure with quadratic scaling, as displayed in
Fig~\ref{fig:Xi_I_pure_cats}.

We note that the \ps Laplacian~$\VEC{\Delta}$ in~(\ref{eq:Xi_Measure}), as well as $\xi$ and ~$\Xi$,
behave as they must: They are \mbox{invariant} under displacements and rotations in \ps, but not
invariant under squeezing~\cite{Bohmann_PRL20}. Additionally, the \ps Laplacian~$\VEC{\Delta}$
in~(\ref{eq:Xi_Measure}) also implements the lowest order, rotation-invariant differential operator
that can be used in a \emph{\sens} quantumness measure, as we show using a `lowest order'-argument
in \ref{subsec:cutoff_xi}.

\section{Operational Value of Quantumness Measures \label{subsec:_operational_value_measures}}

It appears sensible to tie quantumness to opera\-tional values the states might have in technical
applications.  But we would like to caution that this is not straightforward. For instance $\Xi$,
like most quantum measures (see~\ref{subsec:OtherMeasures}), does not change with displacements in
\ps, as is commonly stipulated~\cite{Nimmrichter_Hornberger_PRL13,Bohmann_PRL20}. Yet, adding, say,
a squeezed vacuum state to a large coherent state to increase interferometric reso\-lution can be
challenging. Both, squeezed vacuum and its displaced version carry the same quantumness but
opera\-tionally they are of very different operational value for advanced
interfero\-metry~\cite{Ganapathy_LIGO__PRX23,Ge_Jacobs_Zubairy__NJPQI23}, also
see~\ref{sec:_operational_value_measures_Squeezed_Rotation}.

Fock states of a given quantumness will be useful to suppress photon number noise,
whereas squeezed states of the same quantumness, instead, are better at suppressing phase noise.

Because of such varying contexts, we do not expect a universal connection between quantumness
and its \mbox{opera\-tional} value as a physical resource to exist~\cite{Wood__Quantumness_Quanta23}.

\section{Generalized Smoothened Quantumness Functions\label{sec:Smoothened_xi}}

$\xi$ comes in different flavours~$\xi^{[S,k]}$~\cite{Bohmann_PRL20}
(see~\ref{sec:generalSmoothened}), composed of more or less smooth functions based on more or less
smooth, $S$-parameterised, \ps distributions
${\cal P}_S$~\cite{Cahill_PR69b,Hillery_PR84,Leonhardt_PQE95}
(see~\ref{sec:Appendix_PSDistributions}).
These~$\xi^{[S,k]}$ can be useful for some field correlation measurements~\cite{Bohmann_PRL20}.

Here, we focussed on the special case $\xi^{[S=0,k=\frac{1}{2}]} \equiv \xi$~\cite{Bohmann_PRL20}
based on quantum states' Wigner-$W$ ($W= {\cal P}_0$) and Husimi-$Q$ ($Q= {\cal P}_{-1}$)
-distributions~\cite{Cahill_PR69b,Hillery_PR84,Leonhardt_PQE95}.  That special case of~$\xi$ was
given in Eq.~(\ref{eq:xi_Certification}) above.

The expression of our quantumness measure $\Xi$ of Eq.~(\ref{eq:Xi_Measure}) can correspondingly be
gene\-ralized by basing it on the associated certification functions~$\xi^{[S,k]}$
(\ref{eq:xi_Certification_Sk_Bohmann_appendix}), yielding the associa\-ted gener\-al measures
\begin{flalign}
  \Xi^{[S,k]} = \iint_{-\infty}^{\infty}  dx \; dp
  \;\; \big.\VEC{\Delta} \xi^{[S,k]} (x,p)\Big|_{\xi^{[S,k]} < 0} \; ,
  \label{eq:Xi_Measure_S_k} 
\end{flalign}
where $\Xi$ of Eq.~(\ref{eq:Xi_Measure}) is the special case $\Xi = \Xi^{[S=0,k=1/2]} $.

\begin{figure}[b] \centering
  \includegraphics[width=4.05cm,height=2.5cm]{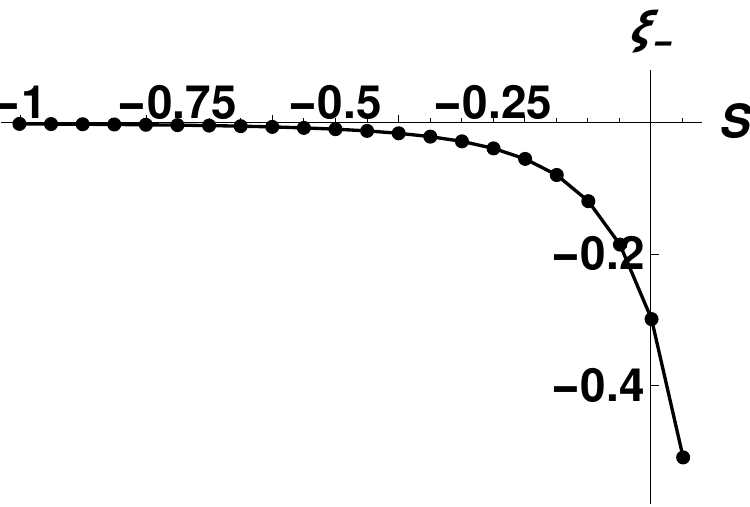}  
  \includegraphics[width=4.05cm,height=2.5cm]{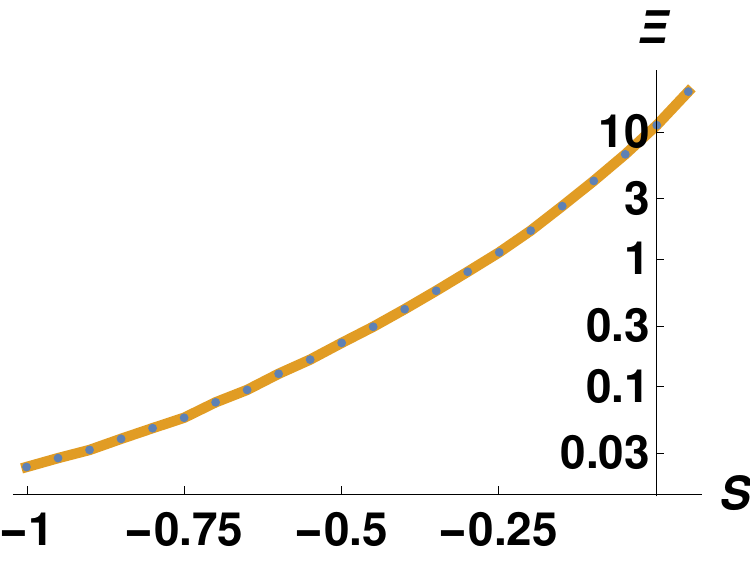}
  \caption{Left panel: values of $\xi_-^{[S,\frac{1}{2}]}$ for a pure cat state~$W_{x_0=3}$.
    Right panel: logarithmic scale plot of~$\Xi^{[S,\frac{1}{2}]}$ for~$W_{x_0=3}$.
    \label{Fig:xi_Xi_S_Cat}}
\end{figure}

These generalized measures are symmetrical in $k$ with respect to its balanced mid-point,
$k=\frac{1}{2}$, see~\ref{sec:generalSmoothened} and Fig.~\ref{Fig:xi_Sk} for more details.

Now we only consider this symmetric case. We highlight that for smoothened input distributions
(parametrized by $S < 0$) the contrast and hence the magnitude of~$\Xi^{[S,\frac{1}{2}]}$ drops, see
Fig.~\ref{Fig:xi_Xi_S_Cat}.

In view of the fact that experimentally the Wigner distribution, $W = {\cal P}_0$, is the best
limiting case of \ps distributions~${\cal P}_S$ that can be directly reconstructed, values $S>0$
should be considered irrelevant for experi\-men\-tal applications. Formally, such reference
distributions, more spiky than $W$~\cite{Cahill_PR69b}, give higher contrasts than~$W$
for~$\Xi^{[S,\frac{1}{2}]}$, see Fig.~\ref{Fig:xi_Xi_S_Cat}. But they can only be generated by the
application of undesirable, noisy deconvolution kernels to experimentally reconstructed
states~\cite{Leonhardt_PQE95}.

Alternatively, the use of \ps distributions~${\cal P}_S$ with $S < 0$ is common in experimental
practice, when noise smoothens distributions~\cite{Leonhardt_PQE95}. When we consider the
performance of the one-parameter family of measures~$\Xi^{[S,\frac{1}{2}]}$ for values $S < 0$ the
use of \ps distributions~${\cal P}_S$, more smeared than ${W = {\cal P}_0}$, should reduce the
contrast in~$\Xi^{[S,\frac{1}{2}]}$. Fig.~\ref{Fig:xi_Xi_S_Cat} confirms this expectation showing a
roughly exponentially drop of the contrast with increasing smearing~$|S|$.

This discussion can be summarized as follows: out of the two-parameter families~ $\Xi^{[S,k]}$ the
measure $\Xi^{[0, \frac{1}{2}]} = \Xi$, of Eq.~(\ref{eq:Xi_Measure}), gives results with the highest
contrast which are in principle achievable with data generated from high quality state
reconstruction measurements~\cite{Lvovsky_Raymer__RMP09}. It is this version of~$\Xi^{[S,k]}$ that
should therefore typically be used, if the experimental reconstruction of the Wigner distribution is
feasible.

\subsection*{Generalization of $\xi$ to a family of sensitive
  quantumness certification functionals\label{subsec:_moreGeneral}}

A small improvement in discrimination between classical and quantum states can be achieved when
using a slightly generalized version of $\xi$, see Ref.~\cite{Ole_25_TowardsQuantumness}. But the
increases in discrimination this can deliver are modest~\cite{Ole_25_TowardsQuantumness} and for
gaussian states, which interest us most, see Sect.~\ref{sec:_apply_experimentatl_data}, $\xi$ is
already faithfully discriminating. We therefore do not include the use of that proposal here.

\section{Conclusions\label{sec:Conclusion}}

We introduce a measure~$\Xi[\state]$ for the quantumness of states,~$\state$, of one-dimensional
continuous-variable systems, represented by their Wigner distributions~$W$.  $\Xi$~is a
function~(\ref{eq:Xi_Measure}) of fixed form allowing for direct like-for-like comparisons of the
quantumness of all states, whether mixed or pure independent of environment, hamiltonian, protocols
or measurements; $\Xi$~is \emph{universal}.

We motivate why we consider $\Xi$ the currently best existing candidate for the quantification of
quantumness that allows for reliable comparisons of the quantumness across all different types of
states, whether pure or mixed.

$\Xi$ is based on the quantumness-certification functional~$\xi$ of Ref.~\cite{Bohmann_PRL20}, from
which $\Xi$ inherits its \emph{discrimi\-na\-tory power}: $\Xi$ is the most sensitive quantification
measure for quantumness we are aware of. For some classes of states it can faithfully discriminate
between classical and nonclassical states~\cite{Ole_25_TowardsQuantumness}.

$\Xi[\state]$ uses the Laplacian in \ps which allows it to sense quantum coherence in phase space
and makes it scale quadra\-tically with an increase in the size of a large \cat state.  For states of
ever increasing quantumness, $\Xi[\state]$ grows without bound.

$\Xi[\state]$ is simultaneously \emph{\sens, \cti}, \emph{monotonic} and \emph{unbounded}.

\section*{Acknowledgments}

This work is partially supported by the National Science and Technology Council of Taiwan (Nos
112-2123-M-007-001, 112-2119-M-008-007, 114-2112-M-007-044-MY3), Office of Naval Research Global,
the International Technology Center Indo-Pacific (ITC IPAC) and Army Research Office, under Contract
No. FA5209-21-P-0158, and the collaborative research program of the Institute for Cosmic Ray
Research (ICRR) at the University of Tokyo.


%

\clearpage


\setcounter{page}{1}
\setcounter{section}{0}
\renewcommand{\thesection}{SI~\arabic{section}}
\renewcommand{\thefigure}{S.~\arabic{figure}}
\setcounter{figure}{0}
\setcounter{equation}{0}
\renewcommand{\theequation}{S.\arabic{equation}}

\onecolumngrid

\begin{center}
  {   \large \bf -- Supporting Information -- \\ \vspace{0.25cm}
    {Universal Quantumness Measure for One-Dimensional Continuous Variable Systems}
}
\\ \vspace{0.25cm}

{Ole Steuernagel\orcidlink{0000-0001-6089-7022}, Hsien-Yi Hsieh\orcidlink{0000-0001-5227-8248}, Yi-Ru Chen\orcidlink{0000-0001-8580-9025}, and Ray-Kuang Lee\orcidlink{0000-0002-7171-7274}}

 \end{center}

\onecolumngrid  

\setcounter{figure}{0}
\vspace{-0.705cm}

\section{Phase Space Probability Distributions\label{sec:Appendix_PSDistributions}}

An electromagnetic radiation field described by creation, $\hat a^\dag$, and annihilation operators,
$\hat a$, can formally be linked to a (mechanical) harmonic oscillator via re-expressing the field
operators in phase space language using operators $\hat x$ and $\hat p$ for position and
momentum. The transformation to the phase space language, which we will use here, is given by the
transformations
$\hat a^\dag(\hat x, \hat p)= \sqrt{\frac{m \omega}{2 \hbar}} (\hat x - {\rm i} \frac{\hat p}{m
  \omega}) $ and
$\hat a(\hat x, \hat p)= \sqrt{\frac{m \omega}{2 \hbar}} (\hat x + {\rm i} \frac{\hat p}{m \omega})
$.
In this work continuous variable system refers to systems with cartesian \ps, unlike, say, systems
on a Bloch sphere~\cite{Yang_PS19}.
Throughout we employ rescaled units, that is, we set $m, \omega$ and $\hbar$ (and also
Boltzmann's constant $k_B$) equal to one.

Note that setting $\omega = 1$ creates a specific harmonic oscillator as a reference
system~\cite{Cahill_PR69b,Scully_Zubairy__Book01}. This is a sensible requirement since also
experimentally the local oscillator, serving as a reference, should evolve at the same fixed
frequency~\cite{Lvovsky_Raymer__RMP09} (generalizations to different mode
structures are possible~\cite{Titulaer_Glauber__PR66}). This implies that only the ground state of this harmonic
oscillator (optical mode) is not squeezed (classical), not the ground states of harmonic oscillators
at other frequencies~\cite{Scully_Zubairy__Book01}. This fixing of the reference system moreover
implies that the only `free' operations~\cite{Tan_Jeong__PRL17,Zhang_Luo__EPJP21}, those which do
not change the nonclassical nature of a state, are displacements and rotations in
\ps~\cite{Bohmann_PRL20}, but squeezing operations are not free.

\subsection{Wigner's Phase Space Distributions\label{subsec:Liouville_W_dist}}

Classical \ps distributions, Liouville distributions,~${f}$, are normalized proba\-bility densities
and therefore obey~${{f}(x,p) \geq 0}$.

In 1932 Wigner introduced the first quantum equiva\-lent of the classical Liouville \ps
distribution~${f}(x,p)$: Wigner's distribution,~$W(x,p)$,~\mbox{\cite{Wigner_PR32,Hillery_PR84}}.
For a one-dimensional continuous system's quantum state described by a density matrix,~$\state$,
$W$~is given by the Fourier transform (an inverse Weyl transform, namely a Wigner
transform~\cite{Daubechies__JMP83}) with respect to the distances~$y$ between its off-diagonal
coherences ${\varrho(x-y,x+y)}$ $ = \langle x-y | \hat \varrho |x+y \rangle$,
\begin{eqnarray}\label{eq:_WignerDistr_Appendix}
  W(x,p) = \frac{1}{\pi \hbar} \int_{-\infty}^{\infty} dy
  \; \varrho(x-y,x+y) \; {\rm e}^{\frac{2 {\rm i}}{\hbar} p y},  
  \quad 
\end{eqnarray}
here, $\hbar=\frac{h}{2\pi}$ is Planck's constant. By construction, $W$ is normalized and nonlocal
(through~$y$). Whereas~$\varrho$ tends to be complex-valued, $W$ is always real-valued but,
generi\-cally, $W$ has nega\-tive values in some regions of
\ps~\cite{Wigner_PR32,Hillery_PR84}. Since~$W$ and~$ \varrho$ are Fourier
transforms~(\ref{eq:_WignerDistr_Appendix}) of each other, they are unitarily equivalent to each other, and
$W$ exists for all states $\state$.  This allows one to describe all aspects of the quantum
system's state and its dynamics using Wigner's \ps representation of quantum
theory~\cite{Zachos_book_05}. Since~$Q$ is a smeared version of $W$ it always exist for all states
as well~\cite{Cahill_PR69b} and both are always well behaved. Therefore, $\xi$ of
Eq.~(\ref{eq:xi_Certification}) also exists for all states and is well behaved.

Unlike the Liouville distribution~${f}$, $W$ is not a probability
distribution~\cite{Mueckenheim__PR86}, this is most clearly seen from the fact that
frequently~$W(x,p) < 0$, which is known to imply that the state $W$ describes is nonclassical.

Implicitly, we adopt the long-standing convention that all probabi\-listic (positive) mixtures of
coherent states are classical~\cite{Glauber__PR63}, but this criterion is based on the
Glauber-Sudarshan distribution ${\cal P}_1(x,p)$, which can be very singular~\cite{Glauber__PR63}
making it difficult to apply it in general, see \ref{subsec:S-parametrized_dist} below.

Wigner's distribution $W$, by contrast, because of relation~(\ref{eq:_WignerDistr_Appendix}), always
exists and is a smooth well behaved function, but there are states which are nonclassical and
yet~${W(x,p) \geq 0}$, that is why we have to use Bohmann and Agudelo's quantumness certification
functional $\xi$ of Eq.~(\ref{eq:xi_Certification}).

Because $W$ and $\state$ are unitarily equivalent and $Q$ is just a smeared version
of~$W$~\cite{Cahill_PR69b,Schleich_01}, see \ref{subsec:S-parametrized_dist}, we can and do use
$\Xi[W]$ and $\Xi[\state]$ interchangeably.

\subsection{Other Phase Space Distributions\label{subsec:S-parametrized_dist}}

There are families of \ps distributions such as the Glauber-Sudarshan distribution $P={\cal P}_1$,
the Wigner distribution~$W={\cal P}_0$ and the Husimi-Q distribution
$ Q = {\cal P}_{-1}$~\cite{Cahill_PR69b}. It is known that they are connected through a smoothing
convolution-identity~\cite{Bohmann_PRL20,Cahill_PR69b}.

Here, since we work in (quantum-mechanical) \ps, we follow the convention of
Ref.~\cite{Leonhardt_PQE95}, which assumes that ${R} \geq {S}$, and connects the various \ps
distributions by
\begin{flalign}
  {\cal P}_{S} (x,p ) = \frac{1}{{R} -{S} } \int_{-\infty}^{\infty} dx' \int_{-\infty}^{\infty} dp'
  \quad {\cal P}_{R}(x',p') \times \frac{1}{\pi} \; \exp \left\{ \frac{-1}{{R} -{S} } \left[ ( x -
      x' )^2 + (p - p')^2 \right] \right\}
  \label{AppendixEq:PDist_sigma_tau} .
\end{flalign}

\begin{figure}[h] \centering
  \includegraphics[width=15.8cm,height=3.8cm]{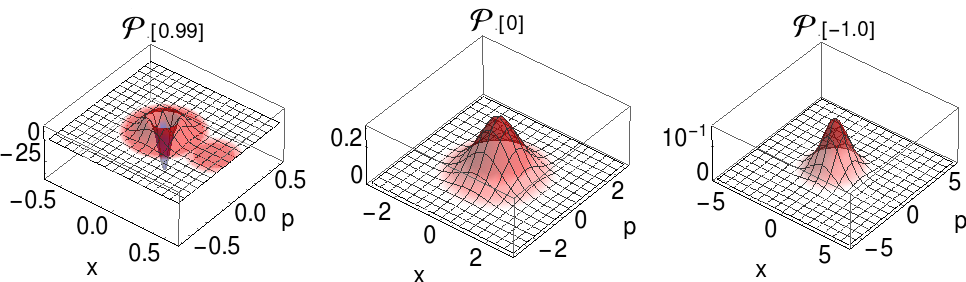}
  \caption{Mixed state with small nonclassical contribution for different smearing parameters $S$
    illustrating that $ {\cal P}[S] (x,p ) $ has lost its telltale negative regions
    when displayed as a Wigner-, $ {\cal P}[0] $, or Q-distribution, $ {\cal P}[-1] $.}
    \label{Fig:_Trends_P_smearing_Appendix}
\end{figure}

To fulfill the commutation relations, Ref.~\cite{Bohmann_PRL20} (which uses complex Glauber state
amplitudes $\alpha^* = ( x - {\rm i} p)/\sqrt{2} $ and $\alpha = ( x + {\rm i} p)/\sqrt{2}$ to
parameterize \ps) uses the expression $\frac{2}{{R} -{S} }$ in exponent and normalization of
Eq.~(\ref{AppendixEq:PDist_sigma_tau}). For us, this leads to an extra factor of `2' in
Eq.~(\ref{eq:xi_Certification}):
\begin{flalign}
 \xi(\alpha) = W(\alpha) - 2 \pi \; Q^2(\alpha) \text{ in
Ref.~\cite{Bohmann_PRL20}}\quad \mapsto \quad \xi(x,p) = W(x,p) - 4 \pi Q^2(x,p), \text{ here.}
\end{flalign}

To transform one into the other type of distribution the appropriate kernels have to be chosen, for
example to convert Glauber-Sudarshan's into Wigner's distribution, start out with
Glauber-Sudarshan's (set ${R} =1$) and apply a smearing Gaussian convolution kernel, based on the
vacuum state, of the form $W_{\squ=1}^{[\mu=1]}(x-x',p-p')$, suitable for transforming into Wigner's
distribution, i.e., set ${S} =0$ in Eq.~(\ref{AppendixEq:PDist_sigma_tau}). Similarly, to map into
Husimi's distribution, choose as above but use the more smeared kernel with ${S} = -1$. Or start out
from Wigner's distribution (set ${R} =0$), now employing the kernel with ${S} =-1$.

Unlike $W = {\cal P}_0$, the so-called Glauber-Sudarshan distribution, ${\cal P}_{S=1}(x,p)$, is so
singular that it remaining posi\-tive throughout \ps certifies that the state must be classical and
if it yields a negative value somewhere in \ps, the represented state is nonclassical:
${\cal P}_{S=1}$ is \emph{discriminating}. In theory, this allows us to certify a state as
quantum. But the closer the value of $S$ in Eq.~(\ref{AppendixEq:PDist_sigma_tau}) gets to `1', the
more difficult it becomes to determine the associated distributions~${\cal P}_{S}$, both
analytically~\cite{Cahill_PR69b} and numerically~\cite{Lvovsky_Raymer__RMP09}.

So, the Glauber-Sudarshan distribution~${\cal P}_{1}$ by itself, in principle, successfully
discriminates between classical and quantum states, but practically speaking, it does not solve our
discrimination problem.  This is why Bohmann and Agudelo's~\cite{Bohmann_PRL20} quantumness certification
function~(\ref{eq:xi_Certification}), based on the Wigner distribution alone, is useful for
practically approaching the discrimination problem.

\clearpage

\section{Necessity for the cutoff $\xi < 0$ in $\Xi$~(\ref{eq:Xi_Measure}) \label{subsec:cutoff_xi}}

The cutoff $\xi<0$ in $\Xi$~(\ref{eq:Xi_Measure}) is somewhat unwelcome since it can make it hard to
derive analy\-tical results. It entails the formal extraction of $\xi<0$. This is difficult because
Bohmann and Agudelo's quantumness certification functional $\xi$ of Eq.~(\ref{eq:xi_Certification})
depends via $Q$ quadratically on the state; moreover, the cutoff is typically very difficult to
treat in analytical calculations, because expressions of the form
$(\xi - |\xi|)/2 = (\xi - \sqrt{\xi^2})/2$, or similar, have to be manipulated.

For example, unlike~$\Xi$, the measure $\cal I$ of Ref.~(\ref{eq:LeeJeong_Measure}) can be written
down explicitly in terms of annihilation and creation operators~\cite{Lee_Jeong__PRL11}; our
measure~$\Xi$ does not seem to permit such an appealing analytical reformulation because of the
cutoff,~$\xi < 0$, in Eq.~(\ref{eq:Xi_Measure}).

But the cutoff is needed: consider the interference pattern of the \cat state. Its
local \mbox{gradients} as well as its higher order derivatives increase with a \cat state's
amplitude~$x_0$. But, integrating out the derivatives over all of the interference pattern does not
work since this will give a zero result, as can be shown after a suitable number of partial
integrations are performed. We are led to integrating over the areas where $\xi < 0$, only. Single
derivatives again give zero. One partial integration (per \ps coordinate $x$ and $p$) shows this.  But
for the next lowest order, the Laplacian $\VEC{\Delta} \xi$, this works, giving us the desired
positive local quantumness contributions to~$\Xi$, if $\xi <0$, see Eq.~(\ref{eq:Xi_Measure}).

This `lowest order'-argument supports our contention that $\Xi$ is the correct measure
(because  we have identified the lowest order derivatives where local
quantumness contributes to $\Xi$).

In other words, using the cutoff implies not only that we exclusively use the `nonclassical parts'
of~$\xi$, thus passing $\xi$'s discrimination power on to $\Xi$. It also is the reason why the
contribution from the Laplacian $\VEC{\Delta} \xi$ does not disappear. The cutoff is unavoidable!

\subsection{Failed attempts to avoid the cutoff $\xi < 0$ in
  $\Xi$~(\ref{eq:Xi_Measure}) \label{subsec:cutoff_xi_avoid}}

The specific case of \cat states, shows that integrating over $\xi$ by itself is insufficient: in
the area where \ps interference happens $\xi \approx W$, see Fig.~\ref{Fig:details_cat}, and
therefore its integral vanishes. Hence, this integral cannot distinguish between coherent
superpositions and a classical mixture of two Glauber states. Also, integrating just over the
negative areas of~$\xi$ does not distinguish between \cat states of different amplitudes, since the
weight of the negative parts of the interference pattern is that of $W$, half of the total
area in which interference occurs~\cite{Kenfack_JOB04}, see Fig.~\ref{Fig:details_cat} and
Eq.~(\ref{eq:Xi_Measure_pure_CatStateApproximation}).

Alternatively, multiplying $\xi$ with some local function before integrating over \ps will typically
break displacement and rotation invariance in \ps. Applying a mapping to~$\xi$, such as $|\xi^m|$ or
similar, and possibly integrating over negative parts of $\xi$ only, seems arbitrary, not well
motivated, and we could not succeed with this type of approach.

One could consider integrating over $\xi$ after convolution with some kernel, perhaps to emphasize
distances in \ps, thus attempting to capture coherent extent of the state.  But this also appears to
be arbitrary, breaking displacement and rotation invariance in \ps, and additionally entails the dangers
of either increasing noise or biases (say, if the kernel is too singular), or smear out features,
thus blunting $\xi$'s ability~\cite{Bohmann_PRL20} to discriminate between classical and
nonclassical states.

\section{\protect{$\Xi$} from line integrals \label{sec:XiFromLineIntegrals}}

Because of the cutoff in the local quantumness contribution to $\Xi$ in Eq.~(\ref{eq:Xi_Measure}),
see \ref{subsec:cutoff_xi} above, the expression for $\Xi$ can be rewritten as line integrals over
all boundaries,~$\partial {\cal B}_j$, of all basins, ${\cal B}_j$, within which $\xi <
0$. According to Green's theorem we have
\begin{flalign}
 \!\! \Xi  = \iint_{-\infty}^{\infty}   \! dx \; dp \; \big.\VEC{\Delta} \xi (x,p)\Big|_{\xi < 0}
= \sum_j \left( \int_{ {\cal B}_j} \! dx \; dp \; \VEC{\nabla} \cdot (\VEC{\nabla}\xi)   \right)  
= \sum_j \int_{\partial {\cal B}_j} \left( dp \; \frac{\partial \xi}{\partial x} - dx \;
  \frac{\partial \xi}{\partial p} \right) \; .
  \label{eq:Xi_MeasureLineIntegral} 
\end{flalign}
We emphasize that the integrand is non-negative, since we integrate along the contour
${\partial {\cal B}_j}$ where $\xi =0$. A line element of the contour $d \VEC{l} = (dx, dp)$ is
perpendicular to $\VEC{\nabla}\xi$, which is pointing `outward'. Hence, a $90^\text{o}$
\emph{clockwise} rotation to $d \VEC{l}_\perp = (dp,-dx)$ aligns these two:
$ \VEC{l}_\perp \cdot \VEC{\nabla}\xi = dp \; \xi_x - dx \; \xi_p = |d \VEC{l}| |\VEC{\nabla}\xi|
\geq 0$.

\subsection{Open Question about Sensitivity\label{sec:Appendix_PS_Sensitivity}}

\begin{figure}[h] \centering
  \includegraphics[width=10cm,height=4.8cm]{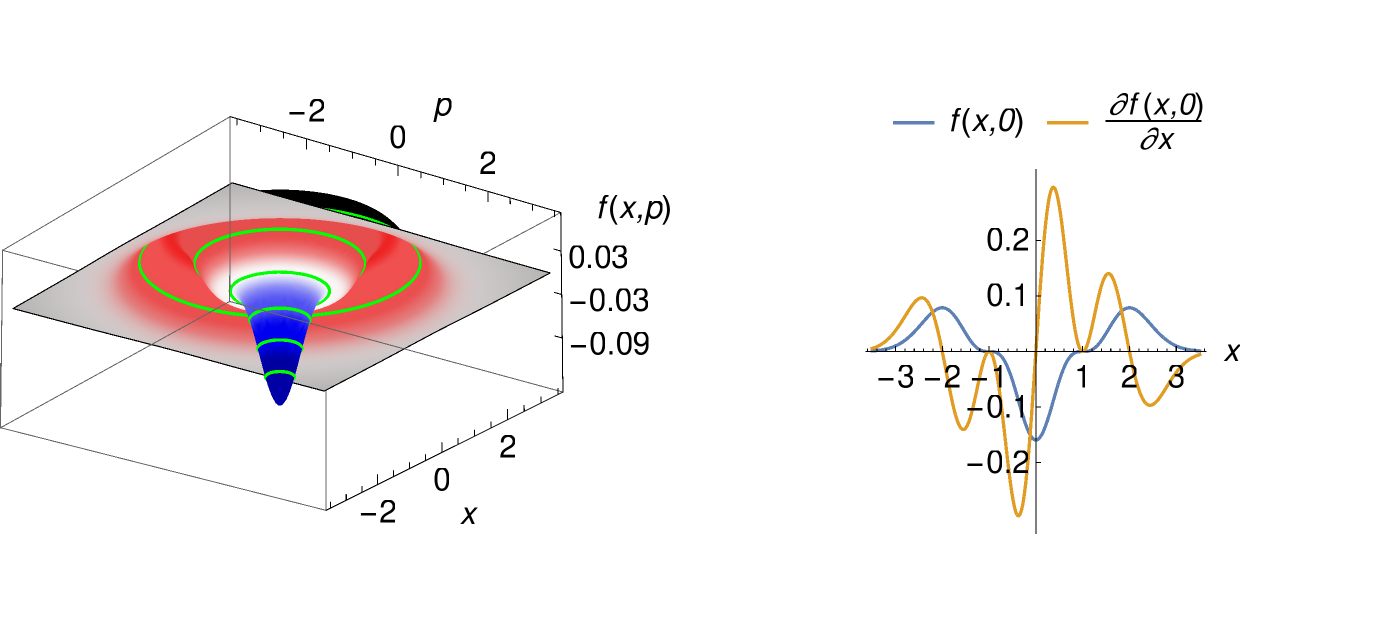}
  \caption{\protect{Function
        $f(x,p) = \frac{\exp({-p^2-x^2})}{2 \pi} \left(p^2+x^2-1\right)^3 $ which has a degenerate
        zero-value contour at radius `1' where also the radial derivative vanishes. Because of this
        double degeneracy, if $f$ were the certification functional $\xi[W]$ of a proper Wigner
        distribution~$W$, the value of the associated quantumness~$\Xi{[W]} $ would equal zero
        despite featuring pronounced negative values. But this $f$ is not associated with the
        certification function~$\xi$ of any viable Wigner distribution.
      }
  \label{Fig:swimmingPoolPlot}}
\end{figure}

There seems to exist a mathematical possibility, that a (simply connected) basin ${\cal B}_j$ with
$\xi < 0$ forms but $\Xi$ does not pick this up. According to Eq.~(\ref{eq:Xi_MeasureLineIntegral}),
this could happen if the zero-value boundary~${{\partial {\cal B}_j}}$ of this basin also has a vanishing
gradient~${\VEC \nabla} \xi$ throughout~${{\partial {\cal B}_j}}$. In this `doubly-degenerate case',
${\VEC \nabla} \xi |_{\xi = 0} =0 $, the integral~(\ref{eq:Xi_Measure}) over ${\cal B}_j$, contributing
to~$\Xi$, would vanish, see Fig.~\ref{Fig:swimmingPoolPlot}.

Although we cannot disprove this case potentially occurring, all that is needed to preclude this
from happening is that the gradient somewhere on the boundary~${\partial {\cal B}_j}$ is
non-zero. Because of the continuous differentiability of~$\xi(x,p)$, this would yield an open
interval on the boundary~${{\partial {\cal B}_j}}$ with non-zero values, thus rendering $\Xi$
positive.

We believe that a `doubly-degenerate' case, in which $\xi$ and its derivatives vanish simultaneously
across an entire boundary~${\partial {\cal B}_j}$, is rare, presumably of measure zero, and thus for practical
purposes irrelevant. Additionally, it is even less plausible that such `double-degeneracy' would
occur on all boundaries ${{\partial {\cal B}_j}}$ of all basins ${\cal B}_j$ of a state~$W$.

Fig.~\ref{Fig:swimmingPoolPlot} shows a general functions with
such `double-degeneracies' but Wigner distributions are special functions, so this case might not
occur.  All cases we checked conform with the expectation that any basin ${\cal B}_j$ of negative
values of $\xi $ is surrounded by a boundary ${{\partial {\cal B}_j}}$ with nonzero slope, and our
attempts to construct a counterexample, using Wigner distributions, failed.

\subsection{Implications for numerical noise\label{sec:Appendix_NumericalNoise}}

$\Xi$'s numerical implementation on a grid of measured values is straightforward since only summing
over the well known approximation for the Laplacian (in the~$x$-direction:
$\frac{\partial^2 \xi}{\partial x^2}$
$\approx [\xi(x+\delta x)-2\xi(x)+\xi(x-\delta x)]/\delta x^2$) for $\xi < 0$ is required (and
likewise for~$p$). These terms form telescoping sums such that, as required, only the boundary terms
at $\xi = 0$ remain uncancelled, see~\ref{sec:XiFromLineIntegrals}. Figs.~\ref{Fig:Trends_Xi}
to~\ref{Fig:Xi_vs_LeeMeasure} were generated from numerical data on grids using this approximation.

Computation of the values of $\Xi$ amounts to the determination of the line integrals
Eq.~(\ref{eq:Xi_MeasureLineIntegral}). This implies that high grid resolution might be needed to
trace the boundaries~$\partial {\cal B}_j$ accurately, otherwise numerical noise can result.

\newpage
\section{Squeezed Vacuum States \label{sec:ImpureSqueezedState}}

For the harmonic oscillator $H = \frac{p^2}{2m} + \frac{k}{2} x^2 $ let us consider its ground
state~$\psi_{|0\rangle}(x) = \left(\frac{1}{\pi} \frac{m \omega}{\hbar} \right)^{1/4} \exp [-\frac{m
  \omega}{\hbar} \frac{x^2}{2}]$. More generally, pure states with normally distributed densities
\begin{eqnarray}
\label{eq:pure_squeezed_psi}
  |\psi(x)|^2 
  = \left(\frac{1}{2 \pi} \frac{2 m \omega}{\hbar} {\frac{1}{V_s}} \right)^{1/2}
  \exp [-\frac{m \omega}{\hbar} {\frac{x^2}{V_s}}]
  = \left(\frac{2}{2 \pi s(V_s)^2} \right)^{1/2} \exp [-\frac{x^2}{s(V_s)^2}] \; \; ,
\end{eqnarray}
where $s(V_s) = \sqrt{V_s \hbar / ( m \omega )}$, describe pure squeezed vacuum states with
variances $\langle \hat x^2 \rangle = \frac{1}{2} \squ^2$.

Specifically, for $V_1 = 1$, $ \psi(x, \squ(1)) $ is the ground state, since
$ |\psi(x, 1)|^2 $$= |\psi_{|0\rangle}(x)|^2 =$ $ \exp [-x^2/(2 \Sigma_0^2)]/\sqrt{2 \pi
  \Sigma_0^2}$, correctly implying the variance is that of the ground state $\langle \hat x^2
\rangle_{|0\rangle} = \frac{1}{2} \squ(1)^2 = \frac{1}{2} \frac{\hbar}{m \omega}$.

In the associated Wigner distribution of a pure squeezed state $W(x,p,V_x,V_p=1/V_x)= \frac{1}{\pi
  \hbar} \exp [{-\frac{p^2 V_x}{ \text{m} \omega \hbar }-\frac{\text{m} x^2 \omega }{{V_x} \hbar }}
] $, (compare expression~(\ref{eq:W_impure_squeezed})) the variances $V_x$ and
$V_p$ can be understood as \emph{relative variances}, such as $V_x =
{\squ^2}/{\squ(1)^2}$.  In this work we set $m=1$, $k=1$, and $\hbar =1$, and so $V_x = 1$ or $V_p =
1$ implies that the fluctuations in $x$ or
$p$ are, respectively, as strong as those of the vacuum. Pure squeezed states are famously Gaussian
and the only pure states with a positive Wigner
distribution~\cite{Hudson_RMP74,Kiesel_PRL11}. Despite this positivity, the nonclassicality
certification function~$\xi$ captures that a pure squeezed vacuum state $W_{\squ} = \exp(-x^2/\squ^2
- \squ^2 p^2)/\pi$ is nonclassical if the squeezing parameter $\squ\neq
1$, since their quantum fluctuations are `too narrow' in one or the other quadrature (below that of
the vacuum state)~\cite{Ole_25_TowardsQuantumness,Wuensche_JOBQSO04}.  Since the nonclassicality
resides in the squeezing fluctuation, \ps displacements and rotations of a squeezed vacuum state do
not change its quantumness.

We mention in passing that for pure squeezed states the uncertainty principle enforces that
$V_x V_p =1$, and that, when including \emph{impure} squeezed vacuum states, this generalizes to
$V_x V_p >1$.

 In this more general case of \emph{impure} squeezed vacuum states the Wigner distributions can be
 written as (see~(\ref{eq:W_impure_squeezed}))
\begin{eqnarray}
  \label{eq:W_impure_squeezed_sLanguage}
  W(x,p,V_x,V_p) = \exp \left[\frac{-x^2}{V_x} + \frac{-p^2}{V_p} \right]
  \left/\left( \pi \sqrt{V_x V_p } \right) \right.
  = W_{\squ}^{[\mu]} = \frac{1}{\sqrt{\mu} \pi} \exp(- \frac{x^2}{\mu V_s} - V_s p^2) \; .
\end{eqnarray}
For definiteness we assume that $V_s = s^2 >1$ such that below-vacuum squeezing always occurs in the
$p$-component and anti-squeezing ($\mu V_s$) [including extra impurity-induced anti-squeezing] in
the $x$-component.  Hence, the
states~(\ref{eq:pure_squeezed_psi})-(\ref{eq:W_impure_squeezed_sLanguage}) are aligned with the
$x$-axis, see Fig.~\ref{Fig:details_squeezed},with respective variances
$\langle \hat p^2 \rangle = \frac{1}{2 V_s} = \frac{1}{2 \squ^2}$ and
$\langle \hat x^2 \rangle = \frac{1}{2} \mu V_s = \frac{1}{2} \mu \squ^2$.

The form~$W_{\squ}^{[\mu]} $ of the impure squeezed state~(\ref{eq:W_impure_squeezed}), or
(\ref{eq:W_impure_squeezed_sLanguage}), is convenient when we consider squeezing on the decibel
scale since it is a relative scale: dB$(\squ^2/\squ(1)^2) = 10 \log_{10}(V_\squ) = $dB$(V_\squ)$.

When considering impure squeezed states $W(x,p,\mu V_\squ,1/V_\squ) $ (\ref{eq:W_impure_squeezed})
we will always assume that $V_\squ > 1$ and that, in accord with the uncertainty principle, the
impurity parameter obeys $\mu \geq 1$. In other words, the states' squeezing in the $p$-direction is
SQ$(\squ^2/\squ(1)^2) = 10 \log_{10}(V_p^{-1}) = 10 \log_{10}(V_\squ)$; and their anti-squeezing in
the $x$-direction is ASQ$(\mu \, \squ^2/\squ(1)^2) = 10 \log_{10}(\mu V_\squ)$.

\begin{figure}[h!]
  \hspace{-0.2cm}
  \begin{minipage}[b]{\columnwidth}
    \; \includegraphics[width=0.35
    \columnwidth,height=3.85cm]{Figures/Squeezed_dB_Scale_ASQ_25_grid_241_241plotXiExperiment_dB_scale_above.pdf}
    \quad
    \includegraphics[width=.22\columnwidth,height=3.85cm]{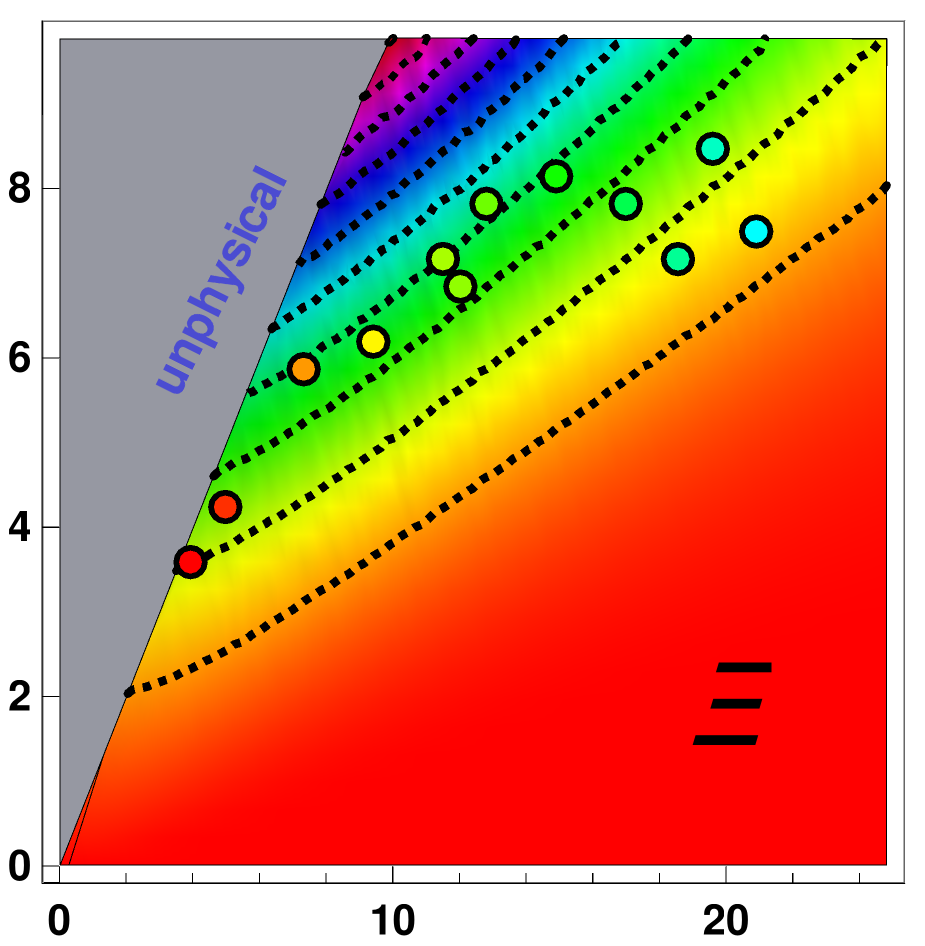}
    \!\!\!\!\!
    \includegraphics[width=.0495\columnwidth,height=3.95cm]{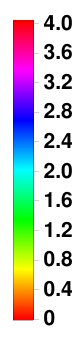}
    \includegraphics[width=.22\columnwidth,height=3.85cm]{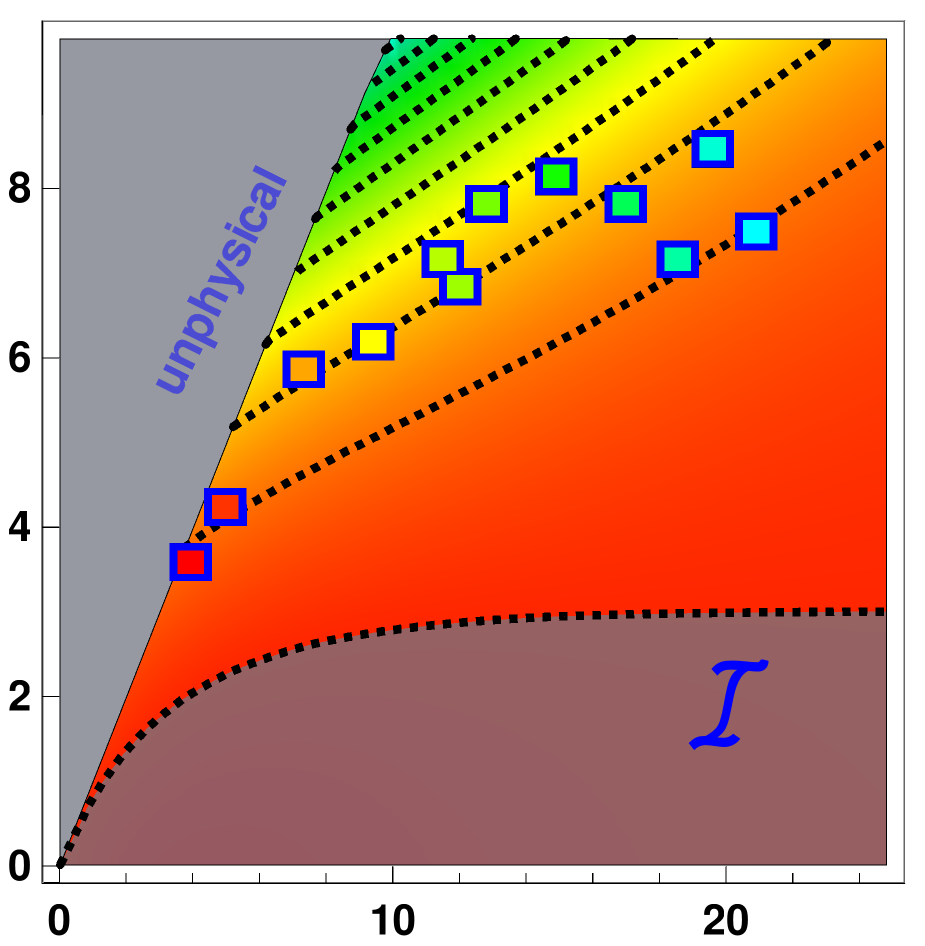}
      \put(-255,93){\rotatebox{90}{  \scriptsize{\textsf{\textbf{\emph{SQ }}}} }}
      \put(-158,-2){\rotatebox{0}{ \scriptsize{\textsf{\textbf{\emph{ASQ}}}} }}
      \put(-115,93){\rotatebox{90}{  \scriptsize{\textsf{\textbf{\emph{SQ }}}} }}
      \put(-19,-2){\rotatebox{0}{ \scriptsize{\textsf{\textbf{\emph{ASQ}}}} }}
      \caption{
      Experimental data (from Fig.~1 of Ref.~\cite{Hsieh__PRL22}) for squeezing and
      anti-squeezing of impure squeezed states $W_\squ^{[\mu]} $~(\ref{eq:W_impure_squeezed}),
      displayed in terms of dB-scales. The gray area signifies unphysical parameters
      violating the uncertainty principle \mbox{($V_x V_p = \mu < 1$).} The data displayed in the Left panel
      are used to determine measures $\Xi$ and $\cal I$  of the associated quantum states
      which are, respectively, overlaid as black circles onto a contour plot of {$\Xi$} (Middle) and as
      blue squares onto that for $\cal I$ (Right). The  color bar (between Middle and Right panels) applies to both contour plots.
      The brown area in the right hand plot represents the area for which for ${\cal I} < 0$, characterizing them as classical
      states, although the states in this area are nonclassical.
      \label{fig:Xi_I_landscape_dB}}
\end{minipage}
\end{figure}

\begin{figure}[h]
  \begin{minipage}[b]{0.97\columnwidth}
    \includegraphics[width=6.5cm/1,height=5cm/1]{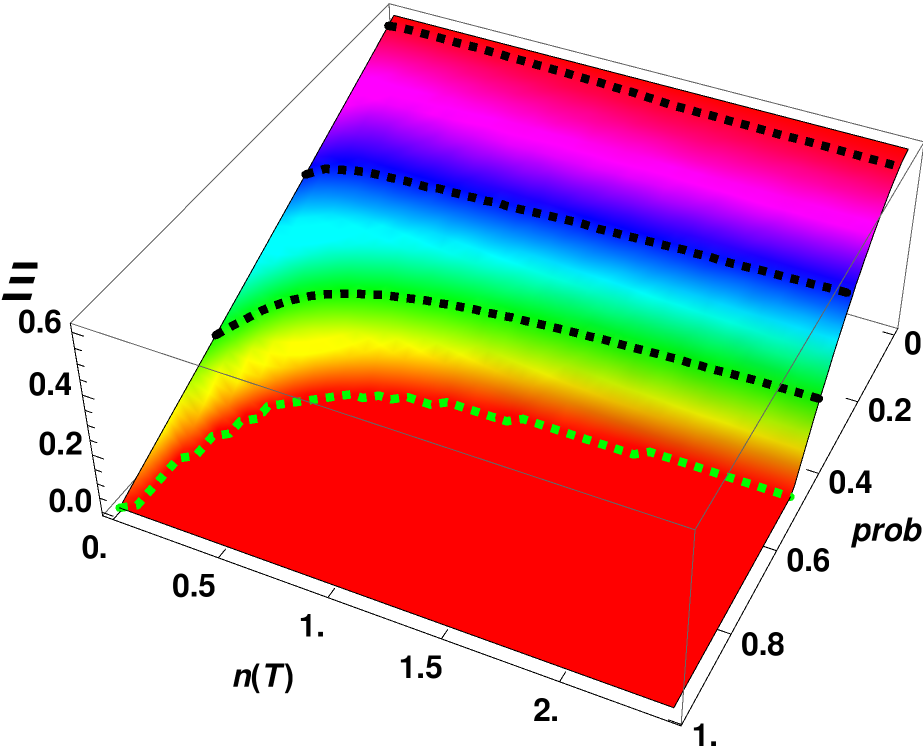}
    \quad
    \includegraphics[width=6.5cm/1,height=5cm]{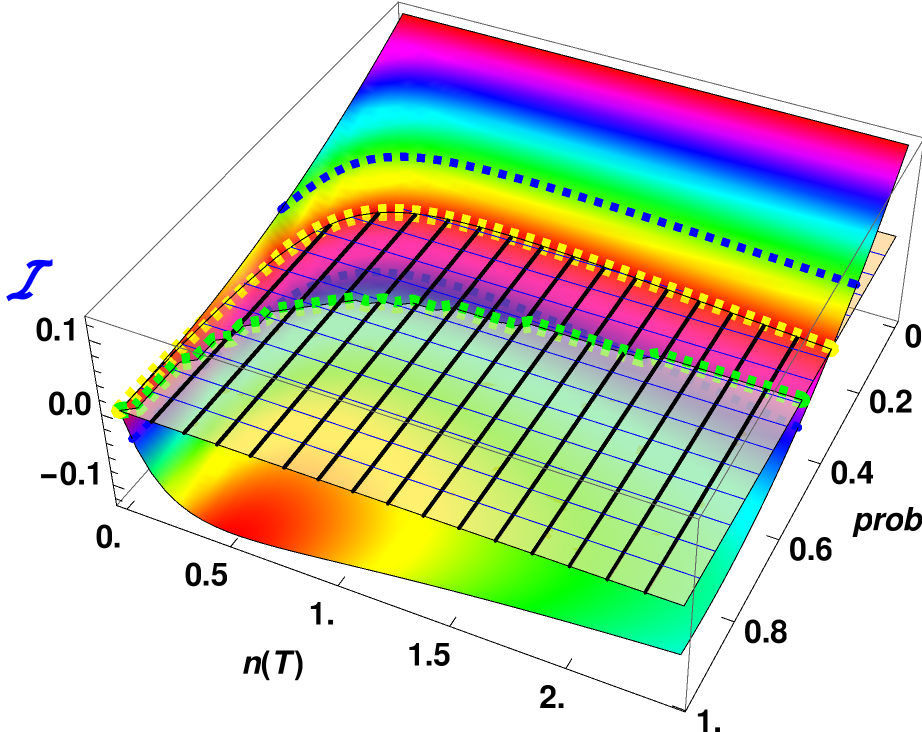}
    \caption{ Quantum measures $\Xi$ and $\cal I$, for mixtures: $W = (1 -
      $prob$) \times W_{\squ = \sqrt{2}}^{[\mu=1]} + $prob$ \times W_{thermal}(n(T))$, where the
      pure squeezed states state (\ref{eq:W_impure_squeezed}) is fixed at 3-dB squeezing
      $V_x =V_s \approx 2$, whereas the thermal state's thermal photon number $n(T)$ ranges from
      zero to 2.5; and the mixing probability weight `prob' runs from zero to one, compare
      Eq.~(\ref{eq:MoreImpurity_LessXi}). Measure $\Xi$
      correctly quantifies states as nonclassical or drops to zero (at green line of red zero-floor) when
      they are classical, as certified by vanishing~$\xi$. Instead,~$\cal I$ dips below zero too
      soon, at the yellow zero contour, rather than at the green zero contour (which has been copied
      over from the $\Xi$-panel on the left). The lack of sensitivity of $\cal I$ behaves
      qualitatively the same as in the scenarios for mixed squeezed and mixed cat states displayed
      in Fig.~\ref{Fig:Xi_vs_LeeMeasure}: ${\cal I} < 0$ can classify states as classical, although
      they have quantum features.
      \label{fig:Xi_Lee_Thermal}}
\end{minipage}
\end{figure}

In terms of decibel, ASQ$(\mu \, \squ^2/\squ(1)^2) = $dB$(\mu) + $dB$(V_\squ) = $dB$(\mu) + $SQ$(V_\squ)$,
$V_\squ = 10^{\text{SQ}/10}$ and $\mu = 10^{(\text{ASQ-SQ})/10}$. This convention is used in
Figs.~\ref{fig:Xi_I_sq_asq},~\ref{fig:XiMeasure_I_sq_asq}, \ref{fig:Xi_I_landscape_dB},
and~\ref{fig:Xi_Lee_Thermal}.

\clearpage

\subsection{Analytical expression for $\Xi$ of Pure Squeezed Vacuum
  States~(\ref{eq:W_impure_squeezed}) \label{sec:PureSqueezedState}}

For pure squeezed states~$W_s^{[\mu=1]}$~(\ref{eq:W_impure_squeezed})
expression~(\ref{eq:Xi_MeasureLineIntegral}) can be determined analytically
yielding~$\Xi$. Insertion of~(\ref{eq:W_impure_squeezed}) into~(\ref{eq:xi_Certification}) and
solving for $\xi=0$ yields parameterized solutions for the loci of vanishing values of $\xi$ located
at $x_{0} (p) = $
$\pm \squ \sqrt{{p^2 \squ ^2+\frac{\squ ^2+1}{\squ^2-1} \left(\log (4)-\log \left(\squ ^2+ 2
        +\frac{1}{\squ ^2}\right)\right)}}$ and $p_{0} (x) = $
$\pm \sqrt{\frac{\left(\squ ^2+1\right) }{\squ ^2 \left(\squ ^2-1\right)}\log \left(\frac{\left(\squ
        ^2+1\right)^2}{4 \squ ^2}\right) +\frac{x^2}{\squ ^4}}$, see highlighting of \mbox{boundary} in
Left Panel of Fig.~\ref{Fig:details_squeezed}.

The basin of vanishing values of $\xi$ is infinite in extent and the values of ${\VEC \Delta} \xi$
die off exponentially, such that only the highlighted boundaries contribute (with a minus sign), see
Left Panel of Fig.~\ref{Fig:details_squeezed}. The second
branch of~(\ref{eq:Xi_MeasureLineIntegral}) gives the contribution\\
$\Xi_{\partial_x \xi} = -\int_{-\infty}^\infty \frac{\partial \xi}{\partial x}|_{x=x_0(p)} dp =
(2^{\frac{1}{\squ ^2-1}} (\squ ^2 + 1)-2^{\frac{\squ ^2}{\squ ^2-1}}) / (\squ ^2+1)^3 \times
\frac{2^{\frac{2 \squ ^2+1}{\squ ^2-1}} \squ ^4 \left(\squ ^2+\frac{1}{\squ
      ^2}+2\right)^{-\frac{2}{\squ ^2-1}}}{\sqrt{\pi } } 
\times U(-\frac{1}{2},0,-\frac{4 \left(\squ ^2+1\right) \log \left(\frac{2 \squ }{\squ
      ^2+1}\right)}{\squ ^2-1})$ where $U$ is the confluent hypergeometric function of the second
kind\\{\href{https://mathworld.wolfram.com/ConfluentHypergeometricFunctionoftheSecondKind.html}{(https://mathworld.wolfram.com/ConfluentHypergeometricFunctionoftheSecondKind.html)}}.

The first branch of~(\ref{eq:Xi_MeasureLineIntegral}), with the contribution
$\Xi_{\partial_p \xi} = -\int_{-\infty}^\infty \frac{\partial \xi}{\partial p}|_{p=p_0(x)} dx$ has a
considerably more complicated form than $\Xi_{\partial_x \xi}$ above, but it is several orders of
magnitude smaller for all values of $\squ >1$ and vanishes quickly the larger $\squ$. Without
any discernible effects on the plot of $\Xi_{\squ}$, presented in Fig.~\ref{Fig:Trends_Xi}, we
therefore ignore it safely. In other words, since there are two branches of $x_0$, we conclude that
a very good approximation is given by $\Xi_\squ = 2 \; \Xi_{\partial_p \xi}$.

\begin{figure}[h] \centering
  \includegraphics[width=6.8cm,height=2.8cm]{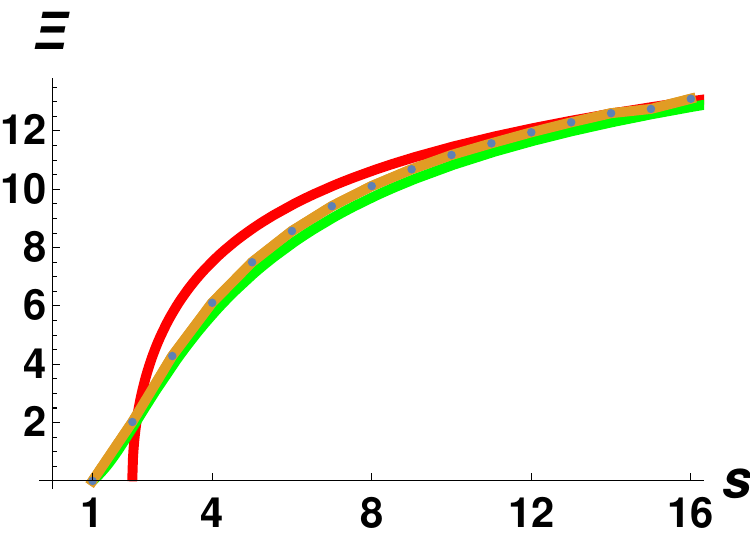}
  \caption{$\Xi$ for pure squeezed,~$W_{\squ}$, brown: numerical data,
    green: exact result~$\Xi_{\partial_x \xi} + \Xi_{\partial_p \xi}$, and
    red: approximation~$\Xi^{[1]}_\squ$.
    \label{Fig:Trends_Xi_squeezed_numerical_approximate}}
\end{figure}

The small discrepancies between numerical results and analytical results visible in
Figs.~\ref{Fig:Trends_Xi} and~\ref{Fig:Trends_Xi_squeezed_numerical_approximate} are due to
numerical noise when implementing the Laplacian of $\Xi_\squ$ on a finite grid, also
see~\ref{sec:Appendix_NumericalNoise}, above.

The approximation $\Xi^{[1]}_\squ \approx 16 ({\frac{1}{\pi} \log (\frac{\squ }{2})})^{\frac{1}{2}}$
is displayed in Fig.~\ref{Fig:Trends_Xi_squeezed_numerical_approximate} and shows that $\Xi_\squ$,
for large values of $\squ$ grows slowly but without bound. 

\clearpage

\section{Smoothing: from~$\xi$ to~$\xi^{[S,k]}$ \label{sec:generalSmoothened}}

In Ref.~\cite{Bohmann_PRL20} the two-parameter families of nonclassicality
certification functions
\begin{flalign}
  { \xi}(S,k) = { {\cal P}}(S) - \frac{\pi(1-S)}{k(1-k)} \; { {\cal P}}(S_k) \; { {\cal P}}(S_{1-k})
  \label{eq:xi_Certification_Sk_Bohmann_appendix} ,
\end{flalign}
with $S_k = 1 - \frac{1-S}{k}$, were introduced. Here the parameter $S\leq 1$ selects the reference
distribution~${ {\cal P}}(x,p;S)$, see Eq.~(\ref{AppendixEq:PDist_sigma_tau}).
Expression~(\ref{eq:xi_Certification_Sk_Bohmann_appendix}) can be useful for some field correlation
measurements~\cite{Bohmann_PRL20}.

The parameter $k \in (0,1)$ linearly interpolates between different exponents in the distributions
${ {\cal P}}(S_k)$ \mbox{${ {\cal P}}(S_{1-k})$}
$ = { {\cal P}}(S_{1-k}) { {\cal P}}(S_{k})$ which are smeared with respect
to~${ {\cal P}}(S)$. Note that, because of the symmetry in $k$ with respect to the
midpoint~$k=\frac{1}{2}$, effectively $k \in (0, 0.5]$.

Also, for vanishing $k$ the limit is $\lim_{k\downarrow 0} S_k= - \infty$ and
$\lim_{k\downarrow 0} S_{1-k} = S$ and we get, according to Eq.~(\ref{AppendixEq:PDist_sigma_tau}),
an infinitely spread-out distribution~${ {\cal P}}(-\infty)$ (giving us an essentially constant
background, which vanishes with increasing smearing values $|S|$) multiplied with~${ {\cal P}}(S)$:
in short, for vanishing $k$, in \ps regions where it assumes negative values,
expression~(\ref{eq:xi_Certification_Sk_Bohmann_appendix}) rises towards zero.

As a function of $k$, negative minimum values ${ \xi}_-(S_k)$ with the grea\-test magnitude, occur
at the balanced mid-point, $k=\frac{1}{2}$, providing the grea\-test contrast when applying $ \xi$,
see Fig.~\ref{Fig:xi_Xi_S_Cat}. This is why here we confine our discussion of $\xi$ to the form of
Eq.~(\ref{eq:xi_Certification}).

Generalizing the expression of our quantumness measure $\Xi$ of Eq.~(\ref{eq:Xi_Measure}), by basing
it on~$\xi^{[S,k]}$, yields
\begin{flalign}
  \Xi^{[S,k]} = \iint_{-\infty}^{\infty}   dx \; dp
  \;\; \big.\VEC{\Delta} \xi^{[S,k]} (x,p)\Big|_{\xi^{[S,k]} < 0} \; .
  \label{eq:Xi_Measure_S_k_Appendix} 
\end{flalign}

Note, for vanishing $k$ the limit is $\lim_{k\downarrow 0} S_k= - \infty$ and
$\lim_{k\downarrow 0} S_{1-k} = S$ and we get, according to Eq.~(\ref{AppendixEq:PDist_sigma_tau}),
an infinitely spread-out distribution~${\cal P}_{-\infty}$ (giving us a constant background)
multiplied with~${\cal P}_S$: in short, for vanishing $k$, expression~(\ref{eq:xi_Certification_Sk_Bohmann_appendix})
rises towards zero everywhere in \ps.

\begin{figure}[h] \centering
  \includegraphics[width=4.05cm,height=2cm]{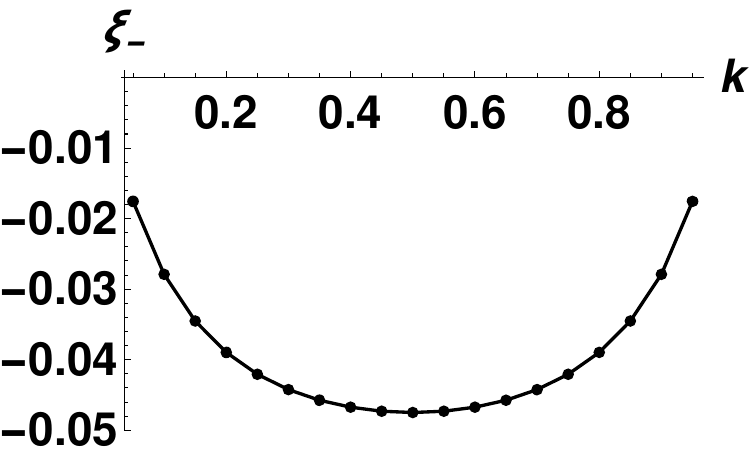}
  \includegraphics[width=4.05cm,height=2cm]{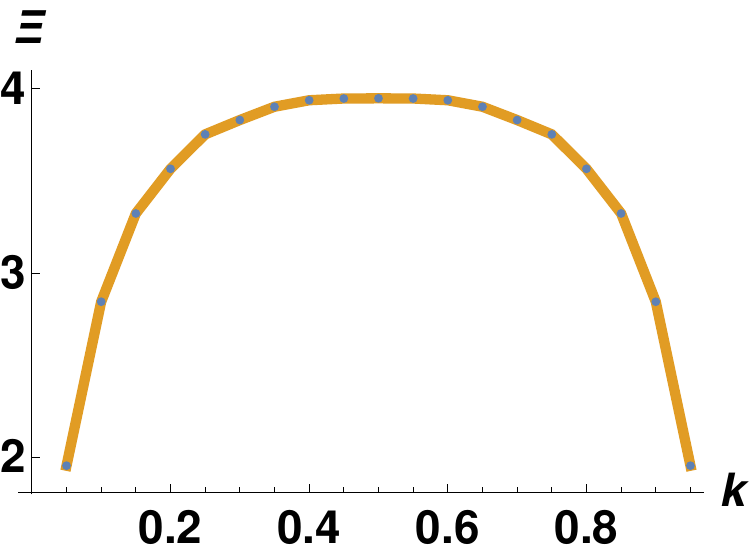}  \\
  \includegraphics[width=4.05cm,height=2cm]{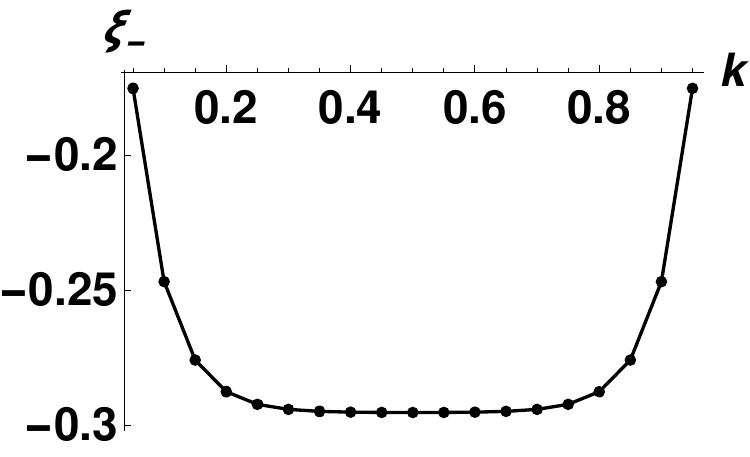}  
  \includegraphics[width=4.05cm,height=2cm]{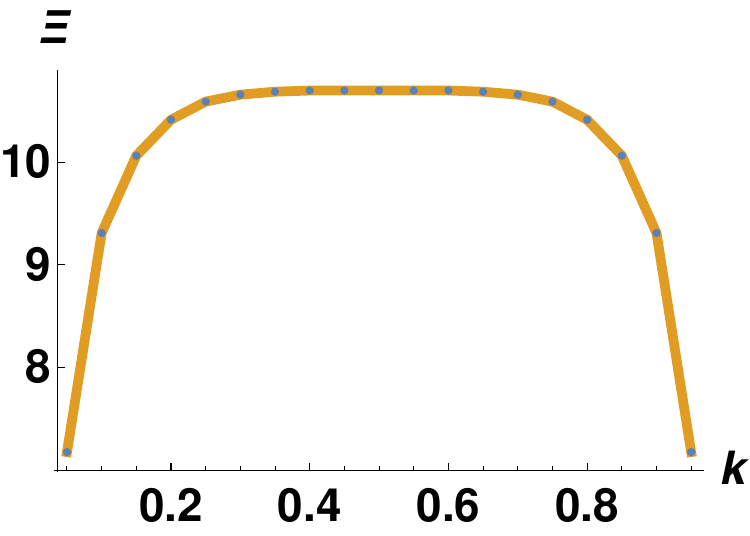}  
  \caption{Top left panel: values of $\xi_-^{[0,k]}$
    for a pure squeezed
    state~$W_{\squ=3}$. Top right panel:~$\Xi^{[0,k]}$ for the same state.
    Bottom left panel: values of $\xi_-^{[0,k]}$
    for a pure cat state~$W_{x_0=3}$. Bottom right panel:~$\Xi^{[0,k]}$ for the same state.
    \label{Fig:xi_Sk}}
\end{figure}

Consequently, we expect to find the lowest minimum values $ \xi_-^{[S,k]}$ at the balanced
mid-point, $k=\frac{1}{2}$. This expectation is confirmed by Fig.~\ref{Fig:xi_Sk} above (but we
could not show this analytically, in the general case).

According to the $k$-symmetry argument just given, we should expect the maxima of $\Xi^{[S,k]}$ to
be reached at the midpoint,~$k=\frac{1}{2}$, as well. Fig.~\ref{Fig:xi_Sk} confirms that this
reason\-ing is correct.

Ideally, the quantumness measure $\Xi$ should be \emph{insensitive} to small variations in $k$,
around the mid-point~$k=\frac{1}{2}$. We find, $\Xi^{[S,k]}$ is indeed insensitive to variations of
$k$ near~$k=\frac{1}{2}$, see Fig.~\ref{Fig:xi_Sk}.

\clearpage

\section{The quantumness certification function~$\xi$ of Eq.~(\ref{eq:xi_Certification}) by itself\\
  cannot be a quantumness measure \label{sec:xi_not_measure}}

The size limit $|W| \leq 1/\pi$ on the maximum amplitude of Wigner's
distribution~\cite{Hillery_PR84} implies that $\xi_-\geq -1/\pi$. With increasing \cat
amplitude,~$x_0$, this limit is rapidly approached from above, see Fig.~\ref{Fig:details_cat}:
$\xi_-$ is not a suitable measure of quantumness for large \cat states.

Widely and coherently spread-out states, such as those studied in~\cite{Ole_JPAMT23}, can show features
of the three families discussed so far. Their dilution of Wigner's distribution $W$, associated with
coherently covering large areas in \ps, typically dimini\-shes their largest (negative) values and
the largest positive values of Husimi's distribution~$Q$. This progressively reduces the global
minimal values $\xi_-$ that $\xi(x,p)$ of Eq.~(\ref{eq:xi_Certification}) can reach, despite the fact
that increased coherent spreading implies increased quantumness: $\xi_-$ cannot serve as a measure.

This conclusion is confirmed for pure Fock states~$|n\rangle$ (i.e., harmonic oscillator
eigenstates). With increasing energy quantum number~$n$ they are increasingly, coherently, spread
out across \ps, see Fig.~\ref{fig:WignerToInfinityPlot}. Yet, their Wigner
distributions~$W_{|n\rangle}$ feature \emph{constant} maximally large negative values,
$\xi(0,0) = \xi_-=-1/\pi$, at the origin when $n$ is odd. For even~$n$, the global mini\-mum
$\xi_-\approx -0.131$ lies in the first annular depression around the origin. Consequently, for Fock
states $\xi_-$ is not a suitable measure of their quantumness, since it \emph{oscillates} between
two fixed values, see Fig.~\ref{Fig:details_fock}.

Similarly, extremely long and narrow (squeezed) posi\-tive ridges or `lines' can form in
\ps~\cite{Ole_JPAMT23}, but as Fig.~\ref{Fig:details_squeezed} shows, $\xi_-$ does not monotonically
drop with increasing squeezing of such features, instead even rising for large squeezing
values~$\squ$.

\section{Operational Value of Quantumness Measures:\\
  Impure Squeezed States and their Sensitivity to Rotation 
  \label{sec:_operational_value_measures_Squeezed_Rotation}}

It appears sensible to tie quantumness to opera\-tional values the states might have in technical
applications.  But we would like to caution that this is not straightforward. For instance $\Xi$,
like most quantum measures (see~\ref{subsec:OtherMeasures}), does not change with displacements in
\ps. Yet, adding, say, a squeezed vacuum state to a large coherent state to increase interferometric
reso\-lution can be very challenging~\cite{Ganapathy_LIGO__PRX23}. Both, squeezed vacuum and its
displaced version carry the same quantumness but opera\-tionally they are of very different value
for advanced interfero\-metry~\cite{Ganapathy_LIGO__PRX23,Ge_Jacobs_Zubairy__NJPQI23}.

Similarly, Fock states of a given quantumness will be useful to suppress photon number noise,
whereas squeezed states of the same quantumness, instead, are better at suppressing phase noise: We
do not expect a \emph{universal} connection between quantumness and its opera\-tional value as an
experimental resource to exist~\cite{Wood__Quantumness_Quanta23}.
\\

To get some intuition on how the quantumness~$\Xi$ compares with the quantification of a specific
application, namely the sensitivity to rotation, we consider the special case of the \emph{impure}
squeezed states of Ref.~\cite{Hsieh__PRL22}. Rotation sensitivity is determined using state-overlap
with itself, $O_+$ (its purity), and upon a \mbox{90-degree} rotation, $O_-$, namely,\\
$O_\pm = 2 \pi \iint dx dp \; W(\theta = 0) \times W(\theta = (1 \mp 1) \times \frac{\pi}{4})$.

We use the associated overlap difference $\Delta O = O_+ - O_-$ and contrast $ C = (O_+ - O_-)/(O_+ + O_-)$.

\begin{figure}[h]
  \begin{minipage}[b]{\columnwidth}
    \includegraphics[width=0.3 \columnwidth,height=3.085cm]{Figures/Squeezed_dB_Scale_ASQ_25_grid_241_241experimentalDataPointsDecibelCombined.pdf}
     \qquad
    \includegraphics[width=0.3 \columnwidth,height=3.085cm]{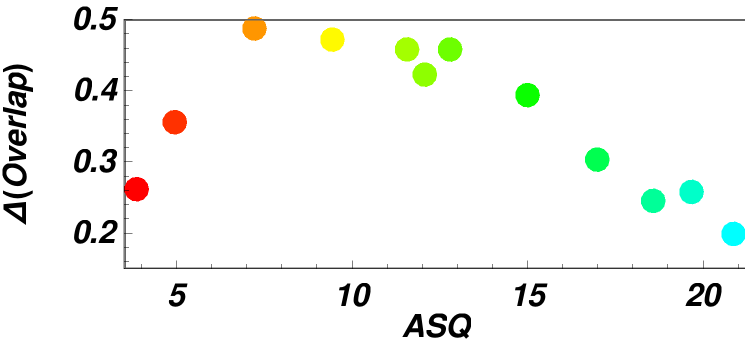}
    \qquad
    \includegraphics[width=0.3 \columnwidth,height=3.085cm]{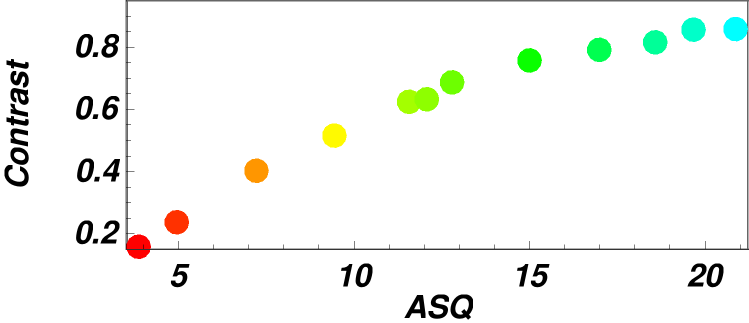}
    \caption{Left: Quantumness values (black circles for measure $\Xi$~(\ref{eq:Xi_Measure}) and
      blue squares for measure ${\cal I}$ of Eq.~(\ref{eq:LeeJeong_Measure})) for the impure
      squeezed states of Ref.~\cite{Hsieh__PRL22} copied over from
      Fig.~\ref{fig:XiMeasure_I_sq_asq}, above (for further details
      see~\ref{sec:ImpureSqueezedState}). Middle: difference  in state overlap $\Delta O$. Right:
      contrast $ C $. In this case quantumness~$\Xi$ behaves similarly to state overlap
      difference $\Delta O$ rather than contrast~$C$.}
      \label{fig:XiMeasure_I_sq_asq_vs_Contrast}
\end{minipage}
\end{figure}

Owing to its circular symmetry, a vacuum state shows zero change in $\Delta O $ whereas a pure
highly squeezed state shows a large drop in state-overlap when rotated.  Similarly, the associated
contrast $ C $ increases with squeezing, see Fig.~\ref{fig:XiMeasure_I_sq_asq_vs_Contrast}.

For impure squeezed states aligned with the $x$-axis and relative variances $V_x$ and $V_p$, see
Eq.~(\ref{eq:W_impure_squeezed_sLanguage}), we have the explicit expressions
$\Delta O = \frac{{V_p}+{V_x}-2 \sqrt{{V_p} {V_x}}}{\sqrt{{V_p} {V_x}}
   ({V_p}+{V_x})} $
and
$C = \frac{V_p+V_x- 2 \sqrt{V_p V_x}}{V_p+V_x+2 \sqrt{V_p V_x}}$.

\clearpage

\section{Other measures\label{subsec:OtherMeasures}}

An overview over measures of states whose quantumness is so large that their quantum effects become
\emph{macroscopic} (as famously discussed by Leggett~\cite{Leggett__JPCM02}) has been given in
Ref.~\cite{Froewis_RMP18}. Some of the approaches discussed in Ref.~\cite{Froewis_RMP18} feature
macroscopic quantumness certification criteria or
witnesses~\cite{Cavalcanti_Reid__PRL06,Cavalcanti_Reid__PRA08}, rather than measures, and are
therefore not discussed further.

We already noticed that measure~\cite{Lee_Jeong__PRL11} is not as sensitive as~$\Xi$, see
Figs.~\ref{Fig:Xi_vs_LeeMeasure},~\ref{fig:Xi_I_landscape_dB} and~\ref{fig:Xi_Lee_Thermal}.

The measure of Ref.~\cite{Kenfack_JOB04}, based on the integral over the negative parts of a Wigner
distribution, shows \emph{non-monotonic} behaviour. Additionally, Ref.~\cite{Kenfack_JOB04} reports
saturation behaviour for larger cat states because such a measure is bounded for
them~\cite{Kenfack_JOB04}, compare saturation of $\xi_-$ in Fig.~\ref{Fig:details_cat}.  On the
other hand, there are Wigner distributions for which the integral over their negative parts is
infinite~\cite{Wlodarz__IJTP03}, see Fig.~\ref{fig:WignerToInfinityPlot}, therefore no rescaling can
be performed to make the approach~\cite{Kenfack_JOB04} a \emph{universal, monotonic} and
\emph{unbounded} measure.

Measures to determine the `nonclassical depth' based on the sensitivity to (Gaussian)
convolution~\cite{Lee__PRA91,Lee__PRA92,Luetkenhaus__PRA95,Marian__PRL02,Malbouisson__PS03} are not
context-independent, in other words, they are not \emph{universal}. They also lack
\emph{monotonicity} (and even continuity as pointed out
by~\cite{Asboth__PRL05}). Refs.~\cite{Lee__PRA91,Lee__PRA92} report different bounds for different
types of states.  Importantly, the reported bounds on different classes of states differ depending
on the type of state.  This, crucially, precludes any attempts to rescale such measures to yield
\emph{monotonic, unbounded} behaviour.

Ref.~\cite{Malbouisson__PS03} explicitly reports lack of \emph{monotonicity}. Ref.~\cite{Asboth__PRL05}
reports \emph{non-monotonic} behaviour for \cat states.

Overlap measure
\cite{Hillery__PRA87,Dodonov_Wuensche__JMO00,Dodonov_Reno__PLA03,Malbouisson__PS03,Boca__PRA09}
entail extremalization, are not context-independent and therefore cannot serve as \emph{universal}
measures. Refs.~\cite{Marian__PRL02,Boca__PRA09} have only been applied to single-mode Gaussian
states.

Refs.~\cite{Hillery__PRA87,Dodonov_Wuensche__JMO00,Dodonov_Reno__PLA03} do not report
results for \cat states.

Approaches quantifying quantumness based on stability against
decoherence~\cite{Nimmrichter_Hornberger_PRL13} and~\cite{Duer_Cirac_PRL02,Shimizu__PRL02} are not
\emph{universal} measures and have not been shown to be sensitive.

The entropic measure presented in~\cite{Asboth__PRL05} does not rise \emph{monotonically} with an
increase in the amplitudes~$x_0$ of \cat states, for `negative cat' states it is bounded by
unity. The entropic measures presented in~\cite{Manfredi__PRE00,vanHerstraeten_Cerf__PRA21} only
apply to states with positive Wigner distributions, they are not \emph{universal}. For some details
see~\cite{Wlodarz__IJTP03}.
The entropic measure of~\cite{Zhang_Luo__EPJP21} is not as sensitive as~$\Xi$, see
Fig.~\ref{Fig:not_discriminating_Zhang_Luo} below.

Measures based on distance in state
space~\cite{Malbouisson__PS03,Marian__PRL02,Aubrun__PRL22,Lee__PRA91,Lee__PRA92,Campos__Pramana16},
or based on measurement operators (such as coherences~\cite{Mandel__PS86,Naseri__PRA21}, photon
number, quadrature amplitude, etc.)  constitute families of measures rather than being
\emph{\cti}. All make comparisons between different states difficult.

The resource theory-based measure of Ref.~\cite{Tan_Jeong__PRL17} reports \emph{non-monotonic} behaviour
for \cat states of varying sizes; the approaches~\cite{Kwon_Jeong_NJP17,Tan_Jeong__PRL20} give
less sensitive measures than $\Xi$.

The measures provided in Refs.~\cite{Froewis_Duerr__NJP12,Yadin_Vedral__PRA16} are
context-dependent, they involve extremalization (are not \emph{universal}) when determining the
quantum Fisher information, and are primarily designed to describe qubit states (discrete multimode
systems). It is unclear how to treat a more irregular state, such as those studied in
Ref.~\cite{Ole_JPAMT23}, with such a measure.

Ref.~\cite{Sekatski__NJP18} also presents a contextual measure involving extremalization.
Ref.~\cite{Laghaout__OC15}, gives several measures all of which are contextual (they are not
\emph{universal}), these measures can be \emph{non-monotonic} and \emph{bounded}.

The `stellar hierarchy' of the zeros of Husimi's distribution does not \emph{monotonically} increase for
\cat states of increasing amplitude~\cite{Chabaud__PRL20}, instead it is infinite for small \cat
states, this is not suitable for a quantumness measure.

The measures~\cite{Bjoerk__JOBQSO04,Korsbakken__PRA07,Marquardt__PRA08} are designed for macroscopic
superpositions and are measurement-operator dependent, hence, they are not \emph{universal}.
\vspace{-0.5cm}
\begin{figure}[h] \centering
  \includegraphics[width=6.8cm,height=3.8cm]{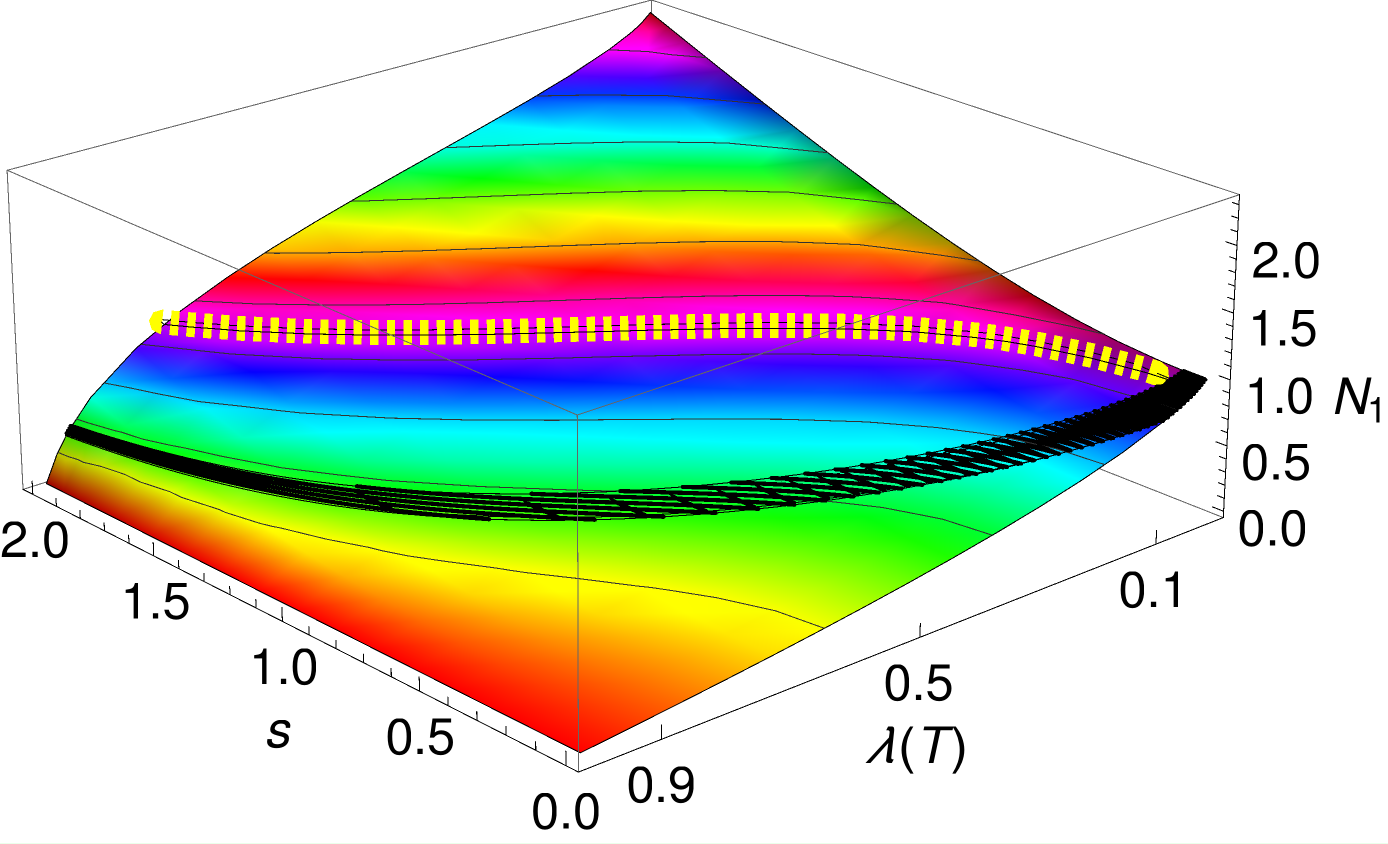}
  \caption{Entropic measure $N_1(s,T)$ of~\cite{Zhang_Luo__EPJP21} for (impure) squeezed thermal
    states as a function of squeezing~$s$ and temperature-dependent Boltzmann-term
    $\lambda(T) = \exp(-\frac{\hbar \omega}{k_B T})$.  $N_1 = 1$ constitutes the (yellow dashed)
    dividing line between classical, $N_1 < 1 $, and quantum states, $N_1 > 1 $ according to Ref.~\cite{Zhang_Luo__EPJP21}. However, 
    the black corridor centers on the area where the vacuum fluctuations of the squeezed thermal
    states equal that of vacuum, truly dividing classical from quantum states~\cite{Wuensche_JOBQSO04}:
    in other words, $N_1$ of Ref.~\cite{Zhang_Luo__EPJP21} is less sensitive than $\Xi$, it mislabels
    states between this corridor and the yellow dashed line as classical although they are nonclassical.
    \label{Fig:not_discriminating_Zhang_Luo}}
\end{figure}

\end{document}